\pdfoutput=1
\RequirePackage{fix-cm}
\documentclass{article}
\usepackage{graphicx}
\usepackage{amssymb}
\usepackage{graphicx, color}

\usepackage{url}
\usepackage{algorithm,algpseudocode,booktabs,amsmath,placeins}
\usepackage{fixltx2e}
\MakeRobust{\Call}

\graphicspath{{../src/plots/}{plots/}{/}}

\newcommand{\ignore}[1]{}

\begin{document}
	
\title{A Novel Feature-Based Approach to Characterize Algorithm
  Performance for the Traveling Salesman Problem
  \thanks{The conference version of this article appeared in the
    proceedings of the \emph{Learning and Intelligent Optimization
      Conference (LION) 2012}~\cite{mersmann2012}.}  }

\author{Olaf~Mersmann$^1$, Bernd~Bischl$^1$, Heike~Trautmann$^1$, \\
Markus~Wagner$^2$, Frank~Neumann$^2$\\
 $^1$ Statistics Faculty, TU Dortmund University, Germany \\
 $^2$ School of Computer Science, The University of Adelaide, Australia\\
  }

\maketitle

\begin{abstract}
  Meta-heuristics are frequently used to tackle NP-hard combinatorial
  optimization problems.  With this paper we contribute to the
  understanding of the success of $2$-opt based local search
  algorithms for solving the traveling salesman problem
  (TSP). Although $2$-opt is widely used in practice, it is hard to
  understand its success from a theoretical perspective. We take a
  statistical approach and examine the features of TSP instances that
  make the problem either hard or easy to solve. As a measure of
  problem difficulty for $2$-opt we use the approximation ratio that
  it achieves on a given instance. Our investigations point out
  important features that make TSP instances hard or easy to be
  approximated by $2$-opt.

  \textbf{keywords}: TSP, $2$-opt, Classification, Feature Selection, MARS
\end{abstract}


\section{Introduction}

For many NP-hard combinatorial optimization problems, meta-heuristic
algorithms such as local search~\cite{HoosS2004}, simulated
annealing~\cite{LaarAar97}, evolutionary algorithms~\cite{EibSmi2007},
and ant colony optimization~\cite{DorigoStuetzleACOBook} have produced
good results. Despite the numerous applications of meta-heuristics to
hard combinatorial optimization problems, it is hard to understand the
success of these algorithms from a theoretical point of view.

Strict mathematical investigations, as in the field of the runtime
analysis of meta-heuristics, allow one to prove when and why such
algorithms are able to solve certain types of problems. This field of
research has gained increasing interest during the last decade and
there are results for a wide range of meta-heuristic approaches such as
simulated annealing~\cite{WegenerSimulated}, evolutionary
algorithms~\cite{DJWoneone}, and ant colony
optimization~\cite{NeumannWittAlgorithmica09}. We refer the reader to
the textbook of Neumann and Witt~\cite{BookNeuWit} for a comprehensive
presentation of this research area.

The rigorous treatment of these algorithms in a strict mathematical
sense is with no doubt desirable, but comes at the expense that one is
usually only able to analyze simplified algorithms on only restricted
classes of problems. With this paper, we follow a different
approach. Our aim is to gain new theoretical insights into the
behavior of meta-heuristics by investigating statistical properties
of hard and easy instances of a given problem for a given algorithm.
This relates to previous work in continuous domains for which the extraction of problem properties that might influence algorithm performance is an
important and current focus of research, denoted as exploratory landscape analysis \cite{MBTPWR11,B2012}. For our investigations on combinatorial meta-heuristics, we choose one of the most famous $NP$-hard
combinatorial optimization problems, namely the traveling salesman
problem (TSP). Given a set of $N$ cities and positive distances
$d_{ij}$ to travel from city $i$ to city $j$, $1 \leq i,j \leq N$ and
$i \ne j$, the task is to compute a tour of minimal traveled distance
that visits each city exactly once and returns to the origin.

In the general case (also known as the asymmetric TSP), the distances
between two cities might even be different, depending on the direction
taken.  Many subclasses of the TSP can be defined depending on the
constraints that the distances between cities have to satisfy. For
example, the distances only have to satisfy the triangle inequality in
the Metric TSP.  The perhaps simplest $NP$-hard subclass of TSP is the
Euclidean TSP where the cities are points in the Euclidean plane and
the distances are the Euclidean distances between them. We will focus
on the Euclidean TSP. It is well known that there is a polynomial time
approximation scheme (PTAS) for this problem~\cite{Arora98}. However,
this algorithm is quite complicated and cumbersome to implement.

A great number of heuristic approaches has been proposed for the
TSP. Often local search methods are preferred in
practice. The most successful algorithms rely on the well-known
$2$-opt operator, which removes two edges from a current tour and
connects the resulting two parts by two other edges such that a
different tour is obtained~\cite{JohnsonG97}. Despite the success of
these algorithms for a wide range of TSP instances, it is still hard
to understand $2$-opt from a theoretical point of view.

In the past, theoretical studies regarding $2$-opt have investigated
the approximation behavior as well as the time to reach a local
optimum.  Chandra et al.~\cite{ChandraKT99} have studied the
worst-case approximation ratio that $2$-opt achieves for different
classes of TSP instances. Furthermore, they investigated the time that
a local search algorithm based on $2$-opt needs to reach a locally
optimal solution.  Englert et al.~\cite{EnglertRV07} have shown that
there are even instances for the Euclidean TSP where a deterministic
local search algorithm based on $2$-opt would take exponential time to
find a local optimal solution. Furthermore, they have derived polynomial
bounds on the expected number of steps until $2$-opt reaches a local
optimum for random Euclidean instances and proved that such a local
optimum gives a good approximation for the Euclidean TSP. These
results also transfer to simple ant colony optimization algorithms as
shown in \cite{AntsTsp10}.  A parameterized analysis of evolutionary
algorithms for the Euclidean TSP using a mutation operator based on
$2$-opt has been recently carried out in~\cite{TSPAAAI12}. These
results show that evolutionary algorithms are provably successful if
the number of cities that lie in the interior of the convex hull of
the given set of $N$ cities is small.

Most previously mentioned investigations have in common that they
either investigate the worst local optimum and compare it to a global
optimal solution or investigate the worst case time that such an
algorithm needs to reach a local optimal solution. Although these
studies provide interesting insights into the structure of TSP
instances they do not give much insights into what is actually
going on in the application of $2$-opt based algorithms. In almost all
cases the results obtained by $2$-opt are much better than the actual
worst-case guarantees given in these papers. This motivates the
studies carried out in this paper, which aim to get further insights
into the search behavior of $2$-opt and to characterize hard and easy
TSP instances for $2$-opt.

In general, meta-learning is a subfield of machine learning, where
learning algorithms are applied to meta-data about experiments.  In
this article, we take a statistical meta-learning approach to gain new
insights into which properties of a TSP instance make it difficult or
easy to solve for $2$-opt. A general overview about how to measure
hardness of instances for combinatorial optimization problems is given
in \cite{SML12}. By analyzing different features of TSP instances and
their correlation we point out how they influence the search behavior
of local search algorithms based on $2$-opt. To generate hard or easy
instances for the TSP we use an evolutionary algorithm approach
similar to the one of \cite{SMHL10}. However, instead of defining
hardness by the number of $2$-opt steps to reach a local optimum, we
define hardness by the approximation ratio that such an algorithm
achieves for a given TSP instance compared to the global optimal
solution. This is motivated by classical algorithmic studies for the
TSP problem in the field of approximation algorithms.

We will consider some minor modifications of the EA used in
\cite{mersmann2012} such that the instances are forced to cover the
whole extent of the underlying plane. Moreover, the rounding procedure
is slightly altered and two different rounding strategies, differing
in the sequence of rounding and mutation, are
investigated. Furthermore, different instance sizes are studied
experimentally.

Having generated instances that lead to a bad or good approximation
ratio, the features of these instances are analyzed and classification
rules are derived, which predict the type of an instance (easy, hard)
based on its feature levels. In addition, instances of moderate
difficulty in between the two extreme classes are generated by
transforming hard instances into easy instances based on convex
combinations of both. We call this procedure
``morphing''. An improved point matching strategy compared to
\cite{mersmann2012} ensures that points move as little as possible
during the transformation. Systematic changes of the feature levels
along this ``path'' are identified and used for a feature based
prediction of the difficulty of a TSP instance for $2$-opt-based local
search algorithms.

The structure of the rest of this article is as follows. In
Section~\ref{sec2}, we give an overview about different TSP solvers,
features to characterize TSP instances and indicators that reflect the
difficulty of an instance for a given solver. Section~\ref{sec3}
introduces an evolutionary algorithm for evolving TSP instances that
are hard or easy to approximate and carries out a feature based
analysis of the hardness of TSP instances. Finally, we finish with
concluding remarks and an outlook on further research perspectives in
Section~\ref{sec:conc}.

\section{Local Search and The Traveling Salesman Problem}
\label{sec2}
As mentioned above, local search algorithms are frequently used to
tackle the TSP. They iteratively improve the current solution by
searching for a better one in its predefined neighborhood. The
algorithm stops when there is no better solution in the given
neighborhood, or if a certain number of iterations has been reached.

Historically, $2$-opt \cite{croes1958} was one of the first successful
algorithms to solve larger TSP instances. It is a local search
algorithm whose neighborhood is defined by the removal of two edges
from the current tour. The resulting two parts of the tour are then
reconnected by two other edges to obtain another complete tour.  A few
years later, this idea was extended to $3$-opt \cite{Lin:1965} where
three connections in a tour are first deleted, and then the best
possible reconnection of the network is taken as a new solution.  Lin
and Kernighan~\cite{LinKin73} extended the underlying idea to more
complex neighborhoods by making the number of performed $2$-opt and
$3$-opt steps adaptive. Nowadays, variants of these seminal algorithms
represent the state-of-the-art in heuristic TSP optimizers.

Among others, memetic algorithms and subpath ejection chain procedures
have shown to be competitive alternatives to the $2$-opt and $3$-opt
based algorithms, with hybrid approaches still being investigated
today.  In the bio-inspired memetic algorithms for the TSP problem
(see \cite{merz2001} for an overview) information about subtours is
combined to form new tours via so-called ``crossover
operators''. Additionally, tours are modified via ``mutation
operators'', to introduce new subtours. The general idea behind the
subpath ejection chain procedures is that in a first step a
dislocation is created that requires further change. In subsequent
steps, the task is to restore the system. It has been shown that the
neighborhoods investigated by the ejection chain procedures form
supersets of those generated by the Lin-Kernighan
heuristic~\cite{journals/dam/Glover96}.

Contrary to the above-mentioned iterative and heuristic algorithms,
Concorde~\cite{journals/informs/ApplegateCDR02} is an exact algorithm
that has been successfully applied to TSP instances with up to 85\,900
vertices. It follows a branch-and-cut scheme~\cite{PadbergRinaldi:91},
embedding the cutting-plane algorithm within a branch-and-bound
search.  The branching steps create a search tree, with the original
problem as the root node.  By traversing the tree it is possible to
establish that the leafs correspond to a set of subproblems that
include every tour for our TSP.

\subsection{Characterization of TSP Instances}\label{subsec:feat}

In general, the theoretical assessment of problem difficulty of a TSP
instance prior to optimization is usually hard if not
impossible. Thus, research has focussed on deriving and extracting
problem properties, which characterize and relate to the hardness of
TSP instances (e.g. \cite{SMHL10,KCHS11,SH11,KM12}). We refer to these
properties as features in the following and provide an overview
subsequently. Features that are based on knowledge of the optimal tour
\cite{SS92,KSW05} cannot be used to characterize an instance a priori
to optimization. They are not relevant in the context of this paper
and thus are not discussed in detail.

A natural and considered feature is the number of cities $N$ of the
given TSP instance \cite{SMHL10,KCHS11,SH11,KM12}. In the following,
the subset of features introduced in \cite{SMHL10,KCHS11,SH11,KM12} which we
incorporated into our study will be detailed. Almost all mentioned
features are included. However, due to the lack of details given in
\cite{KM12}, some of the discussed features had to be omitted.
Furthermore, we present several new features we feel additionally
relevant, i.e. features based on the minimum spanning tree (MST) and
the angle between neighboring cities.

The features can be classified into eight groups which are detailed in
the following. In total, 47 features are considered.
\begin{description}
\item[Distance Features:] One subset of features is based on summary
  statistics of the edge cost distribution. We will use edge cost or
  edge weight synonymously for distance between nodes.
  The lowest, highest, mean and median edge costs are
  considered. The proportion of edges with distances shorter than the
  mean distance, the fraction of distinct distances, i.e. different
  distance levels, and the standard deviation of the distance matrix
  are included as well. The expected tour length for a random tour,
  given by the sum of all edge costs multiplied by $2/(N-1)$,
  completes the list of suitable distance features.

\item[Mode Features:] Additional features \cite{KCHS11} are the number
  of modes of the edge cost distribution and related features such as
  the frequency and quantity of the modes and the mean of the modal
  values. Enhancing the latter approach given in \cite{KCHS11} we
  include a feature for computing the number of modes of the edge cost
  distribution \cite{MBTPWR11}.

\item[Cluster Features:] Smith-Miles et al. \cite{SH11,SMHL10} list
  features that assume that the existence and number of node clusters
  affect the performance of TSP solvers. GDBSCAN \cite{Sander98} as
  recommended in \cite{SMHL10} is used for clustering where
  reachability distances of $0.01$, $0.05$ and $0.1$ are
  chosen. Derived features are the the number of clusters and the mean
  distances to the cluster centroids.

\item[Nearest Neighbor Distance Features:] Uniformity of an instance
  is reflected by the minimum, maximum, mean, median, standard
  deviation and the coefficient of variation of the normalized
  nearest-neighbor distances (nnd) of each node \cite{SH11,SMHL10}.

\item[Centroid Features:] The coordinates of the instance centroid
  together with the minimum, mean and maximum distance of the nodes
  from the centroid are considered.

\item[MST Features:] Statistics are included which are related to the
  depth and the distances of the minimum spanning tree (MST). The
  minimum, mean, median, maximum and the standard deviation of the
  depth and distance values of the MST are completed by the sum of the
  distances on the MST, which we normalize by diving it by the sum of
  all pairwise distances. One reason for considering this feature is
  the MST heuristic which provides an upper bound for the optimal
  tour, i.e. the solution of the MST heuristic is within a factor two
  of optimal \cite{BT97}.

\item[Angle Features:] This feature subset comprises statistics
  regarding the angles between a node and its two nearest neighbor
  nodes, i.e. the minimum, mean, median, maximum and the respective
  standard deviation.

\item[Convex Hull Features:] The area of the convex hull of the
  instance reflects the ``spread'' of the instance in the
  plane. Additionally, the fraction of nodes which define the convex
  hull is computed.
\end{description}

R \cite{RR} source code for the feature computation can be found online\footnote{\label{fn:code} http://www.statistik.tu-dortmund.de/compstat\_supplementary\_material.html}. Note that the features have to be
normalized appropriately in order to allow for a fair comparison of
features across instances of different sizes $N$. Ideally, all
instances should be normalized to the domain $[0,1]^2$ to get rid of
scaling issues. However, the latter will not be an issue for our
experiments as we explicitly generate instances which fill the
$[0,1]^2$ plane.

In order to assess the difficulty of a given TSP instance, we will use
the approximation ratio that an algorithm achieves for this instance
as the optimization accuracy. The approximation ratio is given by the
relative error of the tour length resulting from $2$-opt compared to
the optimal tour length and is a classical measure in the field of
approximation algorithms~\cite{Vaz01}. Based on the approximation
ratio that the $2$-opt algorithm achieves, we will classify TSP
instances either as easy or hard. Afterwards, we will analyze the
features of hard and easy instances.

\def\mat#1{\mathbf #1}
\def\abs#1{\left|#1\right|}

\begin{algorithm}
  \caption{Generate a random TSP instance.}
  \label{alg:randomInstance}
  \begin{algorithmic}
    \Function{randomInstance}{$size$}
    \For{$i = 1 \to size$}
    \State $instance[i, 1] \gets {\cal U}(0, 1)$
    \Comment{Uniform random number between 0 and 1}
    \State $instance[i, 2] \gets {\cal U}(0, 1)$
    \Comment{Uniform random number between 0 and 1}
    \EndFor
    \State \Return $instance$
    \EndFunction
  \end{algorithmic}
\end{algorithm}

\begin{algorithm}
  \caption{EA for evolving problem easy and hard TSP instances}
  \label{alg:EA}
  \begin{algorithmic}
    \Function{EA}{$popSize, instSize, generations, time\_limit, cells, repetitions, type, \newline rnd, mutationParameters$}
    \State $poolSize \gets \lfloor popSize/2 \rfloor$
    \For{$i = 1 \to popSize$}
    \State $population[i] \gets \Call{rescale}{\Call{randomInstance}{instSize}}$
    \State $population[i] \gets \Call{round}{population[i],cells}$
    \If{$rnd$}
    \State $population[i] \gets \Call{normalMutation}{population[i]}$
    \State $population[i] \gets \Call{CutToBoundary}{population[i]}$
    \EndIf
    \EndFor
    \For{$generation = 1 \to generations$}
    \For{$k=1 \to popSize$}
    \State $fitness[k] \gets \Call{computeFitness}{population[k], repetitions}$
    \EndFor
    \State $matingPool \gets \Call{createMatingPool}{poolSize, population, fitness}$
    \State $nextPopulation[1] \gets population[\Call{bestOf}{fitness}]$ \Comment{1-elitism}
    \For{$k=2 \to popSize$}
    \State $parent1 \gets \Call{randomElement}{matingPool}$
    \State $parent2 \gets \Call{randomElement}{matingPool}$
    \State $offspring \gets
      \Call{uniformMutation}{\Call{uniformCrossover}{parent1, parent2}}$
    \If{!$rnd$}
    \State $offspring \gets \Call{normalMutation}{offspring$}
    \EndIf
    \State $offspring \gets \Call{rescale}{offspring}$
    \State $offspring \gets \Call{round}{offspring,cells$}
    \If{rnd}
    \State $offspring \gets \Call{normalMutation}{offspring}$
    \State $offspring \gets \Call{CutToBoundary}{offspring}$
    \EndIf
    \EndFor
    \State $population \gets nextPopulation$
    \If{over time limit $time\_limit$}
    \State \Return $population$
    \EndIf
    \EndFor
    \EndFunction
  \end{algorithmic}
\end{algorithm}

\begin{algorithm}
  \caption{Compute Fitness}
  \label{alg:uniform_crossover}
  \begin{algorithmic}
    \Function{ComputeFitness}{$instance, repetitions$}
    \State $optimalTourLength \gets \Call{concorde}{instance}$
    \For{$j=1 \to repetitions$}
    \State $twoOptTourLengths[j] \gets \Call{twoOpt}{instance}$
    \Comment{Two Opt Tour length}
    \EndFor
    \State \Return $\frac{\Call{mean}{twoOptTourLengths}}{optimalTourLength}$
    \EndFunction
  \end{algorithmic}
\end{algorithm}

\begin{algorithm}
  \caption{Mating pool creation}
  \label{alg:matingPool}
  \begin{algorithmic}
    \Function{createMatingPool}{poolSize, population, fitness}
    \For{$i = 1 \to poolSize$}
    \State $matingPool[i]$
    \State $\gets \Call{betterOf}{
     \Call{randomElement}{population},
      \Call{randomElement}{population}}$
    \EndFor
    \State \Return $matingPool$
    \EndFunction
  \end{algorithmic}
\end{algorithm}

\begin{algorithm}
  \caption{Rescale instance}
  \label{alg:rescale}
  \begin{algorithmic}
    \Function{rescale}{$instance$}
    \State $mins$ $\gets$ \Call{column\_mins}{$instance$}
    \State $maxs$ $\gets$ \Call{column\_maxs}{$instance$}
    \State $scaledPop$ $\gets$ (($instance - mins$)$^{T}$/($maxs - mins$))$^{T}$
    \State \Return $scaledPop$
    \EndFunction
  \end{algorithmic}
\end{algorithm}

\begin{algorithm}[h]
  \caption{Round instance}
  \label{alg:round}
  \begin{algorithmic}
    \Function{round}{$instance,cells$}
    \State $gridRnd     \gets \Call{createGrid}{resolution=cells}$
    \State $instRnd \gets \Call{floor}{instance*cells}/cells$
    \For{$i = 1 \to instSize$}
    \State $instRnd[i,] <- SetToCellCenter(instRnd,gridRnd)$
    \EndFor
    \State \Return $instRnd$
    \EndFunction
  \end{algorithmic}
\end{algorithm}

\section{Analysis of TSP Problem Difficulty}\label{sec3}

In this section, we analyze easy and hard TSP instances. We start by
describing the evolutionary algorithm that we used to generate these
instances. Later on, we characterize them using different features
which we calculated and analyzed to determine which features make a
TSP instance difficult or easy to solve for $2$-opt.

\subsection{EA-Based Generation of Easy and Hard TSP Instances}
Our aim is to identify the features that are crucial for predicting
the hardness of instances for the $2$-opt heuristic. For this a
representative set of instances is required which contains instances
of varying degrees of difficulty. It turned out that the construction
of such a set is a nontrivial task. The generation of instances in a
random manner did not provide a sufficient spread with respect to the
instance hardness. The same is true for moderately sized instances
contained in the TSPLIB, i.e. lower than 1000 nodes, for which, in
addition, the number of instances is not high enough to provide an
adequate basis for our analysis. Higher instance sizes were excluded
due to the large computational effort required for their analysis,
especially the computation of the optimal tours.

Therefore, two sets of instances are constructed in the $[0,1]^2$-plane,
which focus on reflecting the extreme levels of difficulty.  An
evolutionary algorithm (EA) is used for this purpose (see
Algs.~\ref{alg:randomInstance} - \ref{alg:round} for a description),
which can be parameterized such that its aim is to evolve instances
that are either as easy or as hard as possible for a given instance
size. The approach is conceptually similar to \cite{SMHL10} but
focusses on approximation quality rather than on the number of swaps
as in our view this indicator more adequately reflects problem
hardness. In addition, the EA concept consists of a different mutation
strategy. Initial studies showed that a second mutation strategy was
necessary. ``Local mutation'' was achieved by adding a small normal
perturbation to the location (normalMutation). ``Global mutation'' was
performed by replacing each coordinate of the city with a new uniform
random value (uniformMutation). This later step was performed with a
very low probability. The two sequential mutation strategies together
enable small local as well as global structural changes of the
offspring resulting from the crossover operation. All parameters are
given at the end of this section.

\begin{figure}
  \centering
  \includegraphics[width=0.35\textwidth]{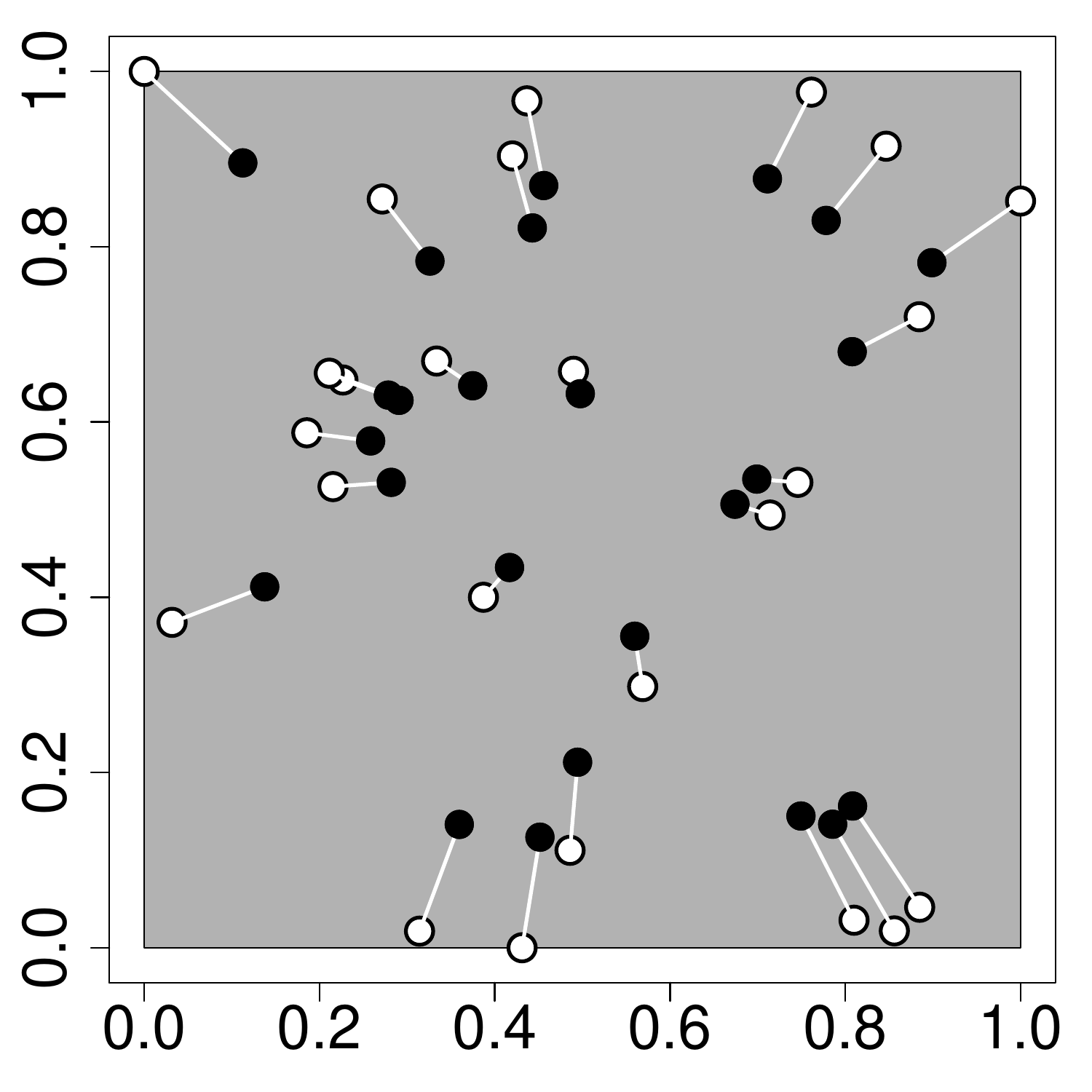}
  \includegraphics[width=0.35\textwidth]{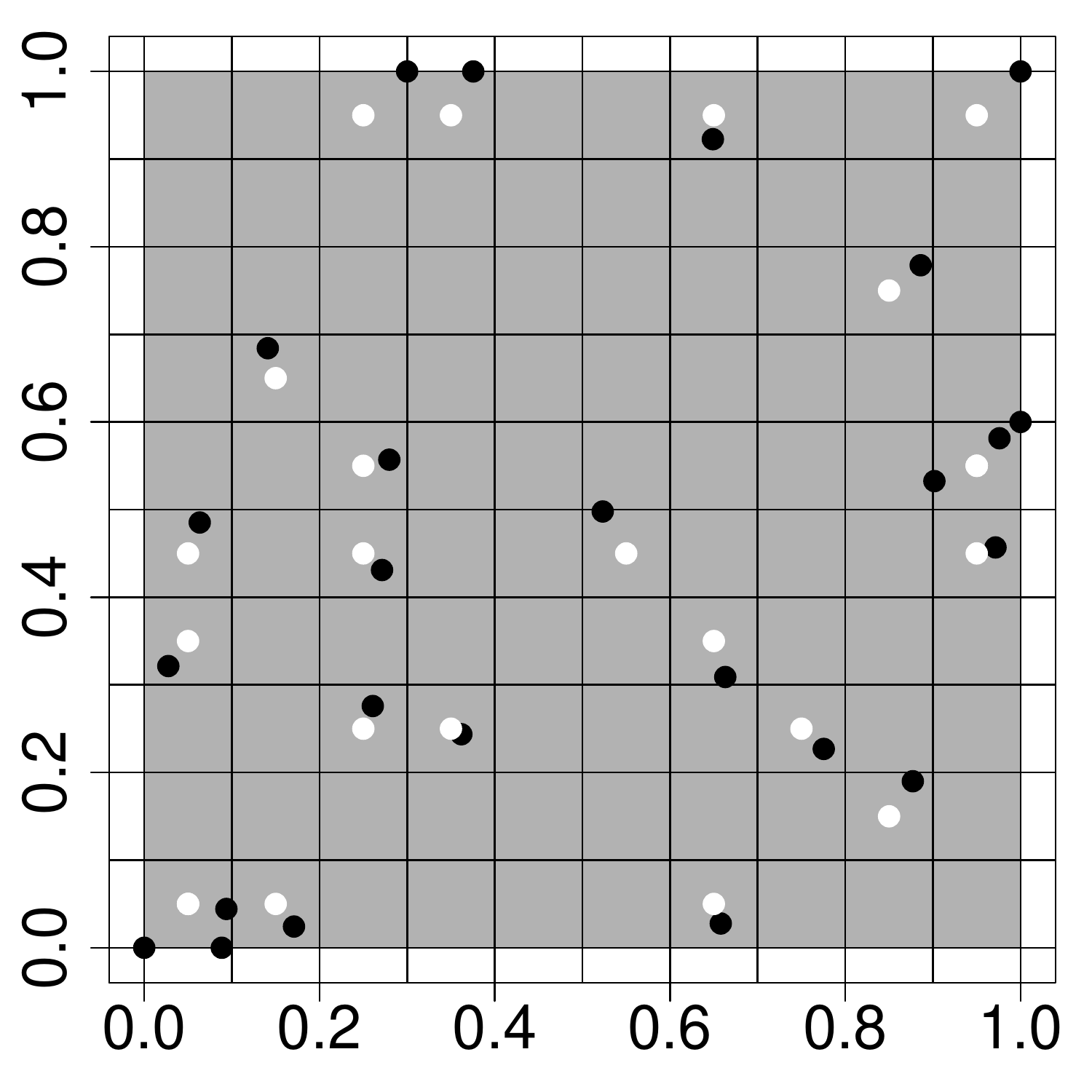}
  \caption{Examples. Left: Rescaling of an instance of size 25. The
    original instance is reflected by black dots. Right: Rounding of
    an instance of size 25 to grid cell centers. The rounded instance
    is visualized by white dots.}
  \label{fig:exRescRound}
\end{figure}

In contrast to our previous work in \cite{mersmann2012}, a rescaling
of the generated instances ensures the complete coverage of $[0,1]^2$
in that the minimum and maximum coordinates are placed on the boundary
of the instance space (see Fig.~\ref{fig:exRescRound}). Therefore the
area covered will not vary as much as in our previous work and
instances become comparable in this regard.

In addition, two different rounding schemes are investigated which
differ in the sequence of the rounding and normal mutation step. In
the first case rounding is applied after both mutation steps are
complete ($!rnd$, denoted as $nrnd$ in the following). After rescaling
of the generated instance the points are rounded to force the cities
to lie on a predefined grid. This is advantageous for some features
which incorporate the proportion of distinct distances. Secondly, we
consider normal mutation after the sequence of uniform mutation,
rescaling and rounding ($rnd$). This strategy results in instances
which resemble a grid structure but also include slight perturbations
of the latter as it for instance occurs in circuit board problems. The
rounding scheme conceptually differs from rounding to a predefined
number of digits as previously considered in that in the current
approach the points are rounded to the center of the grid cell they
are placed in (see Fig.~\ref{fig:exRescRound}). By this means the
probability that cities are located outside the boundary after normal
mutation of the rounded points is very low. In these cases points are cut to the boundary of the plane.

The fitness function to be optimized is chosen as the approximation
quality of $2$-opt, estimated by the arithmetic mean of the tour
lengths of a fixed number of $2$-opt runs, on a given instance divided
by the optimal tour length which is calculated using
Concorde~\cite{journals/informs/ApplegateCDR02}. In general other
summary statistics instead of the arithmetic mean could be used as
well such as the maximum, minimum or median approximation quality
achieved. Note that randomness is only induced by varying the initial
tour whereas the $2$-opt algorithm is deterministic in always choosing
the edge replacement resulting in the highest reduction of the current
tour length.  Depending on the type of instance that is desired, the
\textsc{betterOf} and \textsc{bestOf} operators are either chosen to
minimize or maximize the approximation quality.

We use a $1$-elitism strategy such that only the individual with the
current best fitness value survives and will be contained in the next
population. The rest of the population is obtained by choosing two
parents from the mating pool, applying uniform crossover, uniform and
normal mutation, rescaling and rounding in the appropriate order and
adding the offspring to the population. This procedure is repeated
until the population size is reached.

In the experiments, 100 instances each for the two instance classes
(easy, hard) with fixed instance sizes of 25, 50 and 100 are
generated. The remaining parameters are set as follows:
$\mathit{popSize} = 30$, $generations=5000$, $time\_limit=24h$,
$\mathit{uniformMutationRate} = 0.001$, $\mathit{normalMutationRate} =
0.01$, $cells = 100$, and the standard deviation of the normal
distribution used in the \textit{normalMutation} step equals
$\mathit{normalMutationSd} = 0.025$. The parameter levels were chosen
based on initial experiments. The number of $2$-opt repetitions for
calculating the approximation quality is set to 500. Again this was a
trade-off between evaluation speed and the noise level of the fitness
function.

\subsection{Characterization of the Generated Instances}
\label{sec:charInst}

\begin{table}
\begin{tabular}{rccrr}
  \toprule
Size & Class & Type & Mean Approximation Quality & Mean \# of generations \\
  \midrule
 25 & easy & nrnd & 1.00 & 5000.00 \\
   25 & easy & rnd & 1.00 & 5000.00 \\
   25 & hard & nrnd & 1.13 & 5000.00 \\
   25 & hard & rnd & 1.13 & 5000.00 \\\midrule
   50 & easy & nrnd & 1.00 & 3991.42 \\
   50 & easy & rnd & 1.00 & 3295.77 \\
   50 & hard & nrnd & 1.16 & 5000.00 \\
   50 & hard & rnd & 1.16 & 5000.00 \\\midrule
  100 & easy & nrnd & 1.03 & 454.93 \\
  100 & easy & rnd & 1.03 & 453.74 \\
  100 & hard & nrnd & 1.18 & 1194.59 \\
  100 & hard & rnd & 1.18 & 1204.87 \\
   \bottomrule
\end{tabular}

  \caption{Overview of generated instances: Mean approximation
    quality and mean number of EA generations within the time limit.}
  \label{tab:qualgen}
\end{table}

Table~\ref{tab:qualgen} gives an overview of the instance generation
process, i.e. the mean approximation qualities and average number of
generations the EA managed to execute within the time limit. For all
instance sizes, a sufficiently high performance discrepancy between
the two evolved sets of hard and easy instances is generated while the
absolute performance difference increases along with the instance
size. For instance sizes 25 and 50, the EA even evolves instances
which $2$-opt nearly manages to solve to optimality on average.

\begin{figure}
  \centering
  \includegraphics{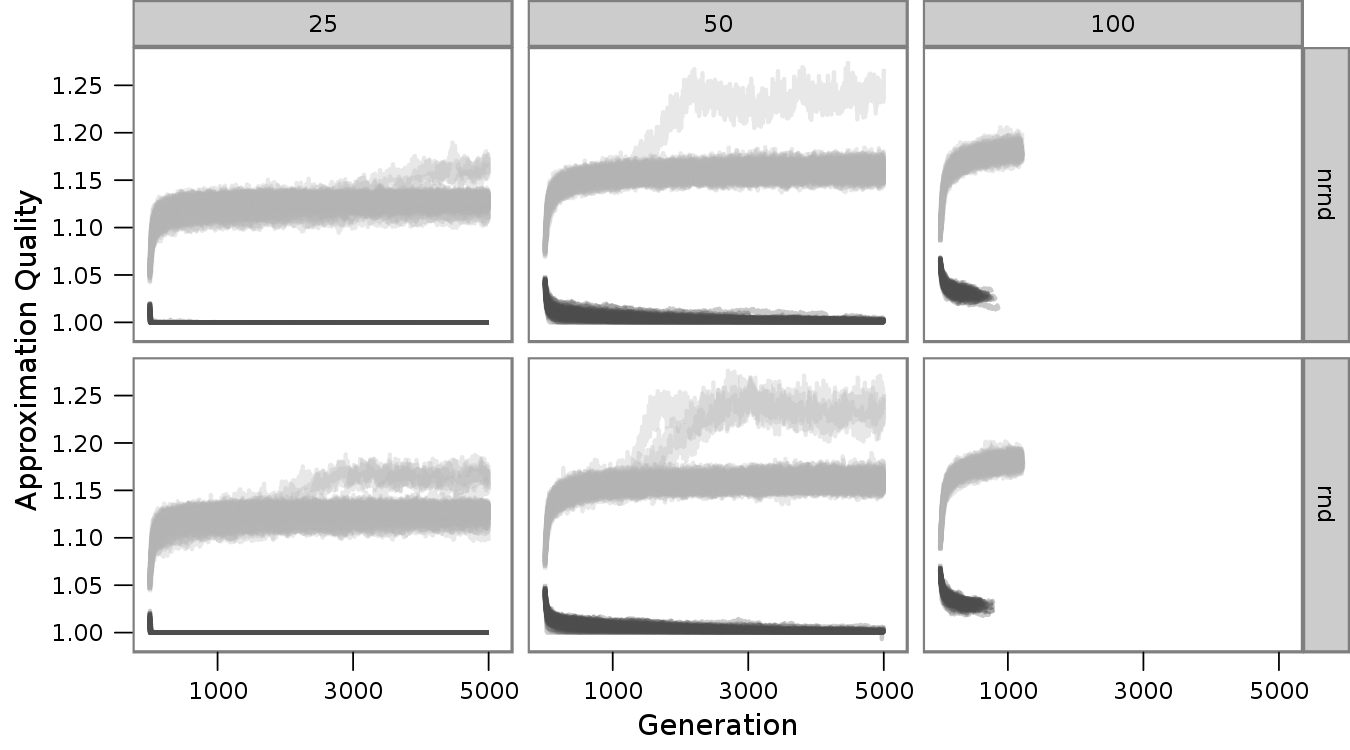}
  \caption{EA fitness in the course of the generations for all
    executed runs.}
  \label{fig:eafitness}
\end{figure}

The average number of generations executed by the EA within the
computing budget reflects the rising computational complexity when
enlarging the instance size. While for the smallest instance size the
maximum number of generations was reached, this amount decreases
substantially for the higher instance sizes. Interestingly, the EA
requires much higher computation times for generating the easy
instances than it is the case for the hard ones. While the number of
swaps carried out by $2$-opt slightly increases in this situation (see
discussion below), particularly problem hardness seems to increase for
Concorde as the algorithm takes up much higher computation times than
for the instances which are hard to approximate for $2$-opt. No
differences can be detected concerning the sequence of rounding and
mutation.

\begin{figure}
  \centering
  \includegraphics{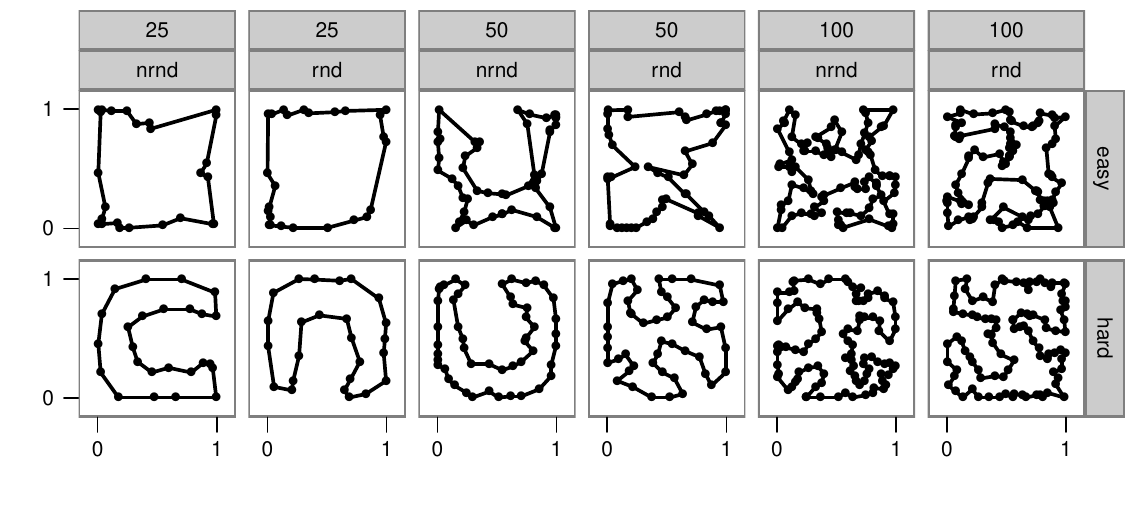}
  \caption{Examples of the evolved instances of both types (easy,
    hard) including the optimal tours computed by Concorde for
    different instance sizes and rounding strategies (round before
    normal mutation (rnd), round after normal mutation (nrnd)).}
  \label{fig:extour}
\end{figure}

\begin{figure}
  \centering
  \includegraphics[width=\textwidth]{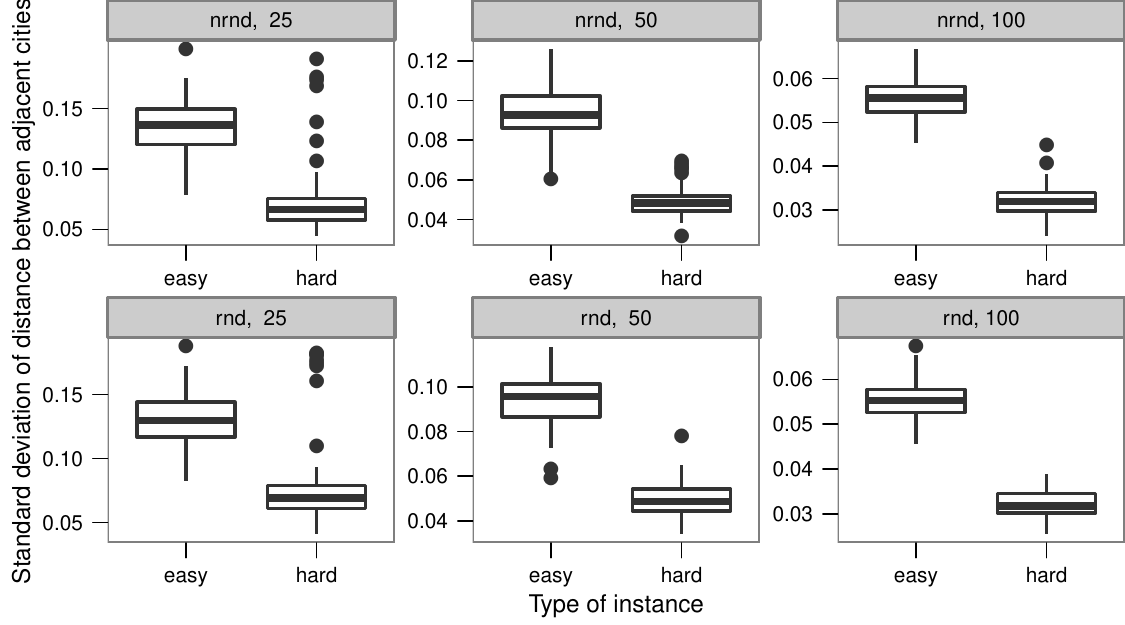}
  \caption{Boxplots of the standard deviations of the tour length legs
    of the optimal tour, both for the evolved easy and hard
    instances.}
  \label{fig:sdtour}
\end{figure}

\begin{figure}
  \centering
  \includegraphics{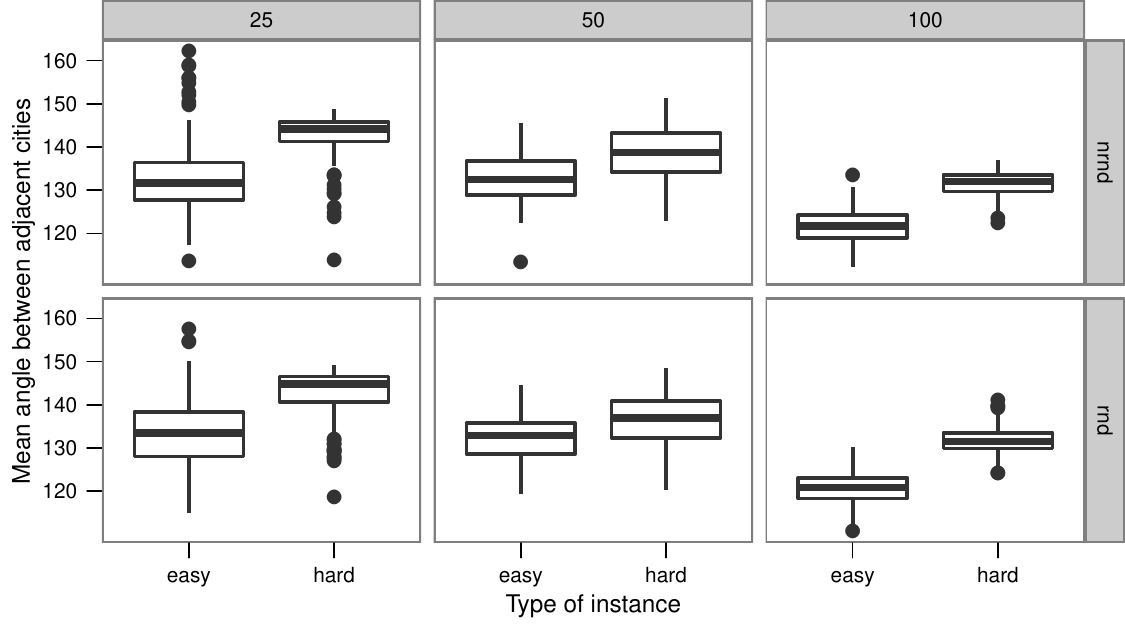}
  \includegraphics{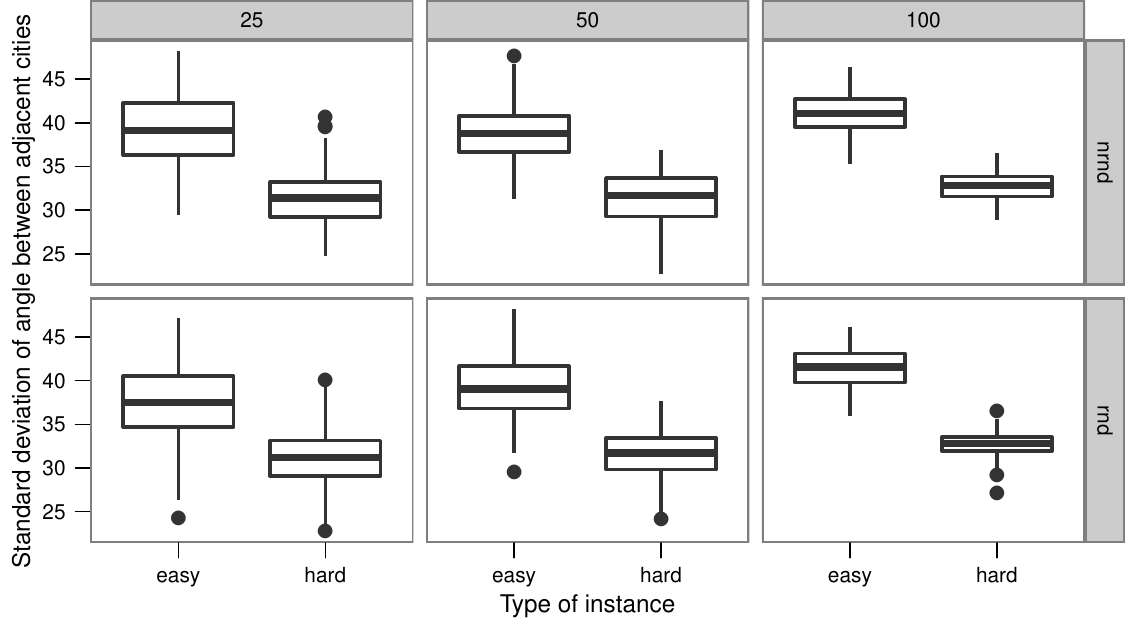}
  \caption{Boxplots of the mean (top) and standard deviations (bottom)
    of the angle between adjacent cities on the optimal tour.}
  \label{fig:anglestour}
\end{figure}

In Figure~\ref{fig:extour} exemplary EA instances of both classes are
shown for the different instance sizes and rounding schemes together
with the corresponding optimal tours computed by Concorde. The main
visual observations can be summarized as follows:
\begin{itemize}
\item The distances of the cities on the optimal tour appear to be
  more uniform for the hard instances than it is the case for the easy
  ones. This is supported by Figure~\ref{fig:sdtour} that shows
  boxplots of the standard deviations of the edge weights on the
  optimal tour. There we see that respective standard deviations of
  the easy instances are roughly twice as high than for the hard
  instances for instance sizes of 100 which increases to a factor of
  three for the smallest instance size. Related to this context it is
  observable that the easy instances tend to consist of many small
  clusters of cities whereas this is not the case for the hard
  instances up to the same extent.
\item Visually, the fraction of highly pointed angles within the class
  of easy instances exceeds the respective proportion within the class
  of hard instances. Fig.~\ref{fig:anglestour} shows mean angles
  between neighboring points on the optimal tour and the corresponding
  standard deviations. The mean angles are significantly smaller for
  the easy instances than within the class of hard ones while the
  opposite is true for the respective standard deviations.
\item The instance shapes for the smallest instance size structurally
  differ from the respective ones regarding the higher instance
  sizes. This is especially the case for the easy instances which
  exhibit an almost circular structure. Consequentially, the area
  within the convex hull enclosed by the points is much higher for high instance sizes than for smaller ones.
\item U-shaped instances are prevalent within the class of generated
  hard instances while the respective frequency increases with
  decreasing instance size.
\item No significant structural differences between the considered
  rounding schemes can be observed. Note that both the U-shaped and
  X-shaped hard instance for the instance size 50 represent
  interchangeable exemplary instances for both $rnd$ and $nrnd$.
\end{itemize}

\begin{figure}
  \centering
  \includegraphics{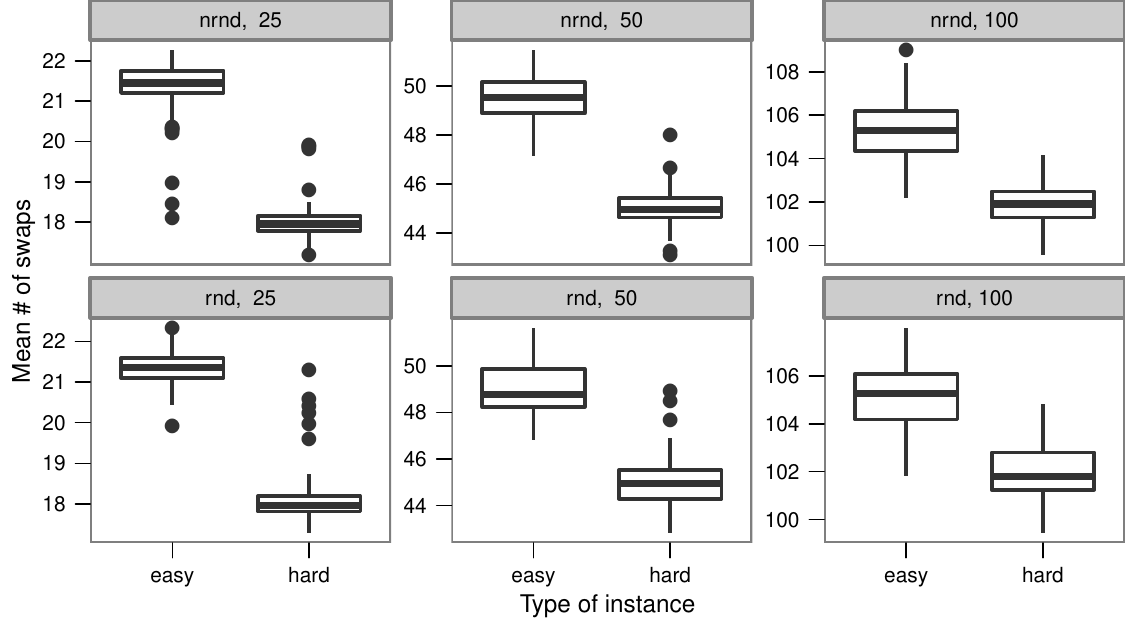}
  \includegraphics{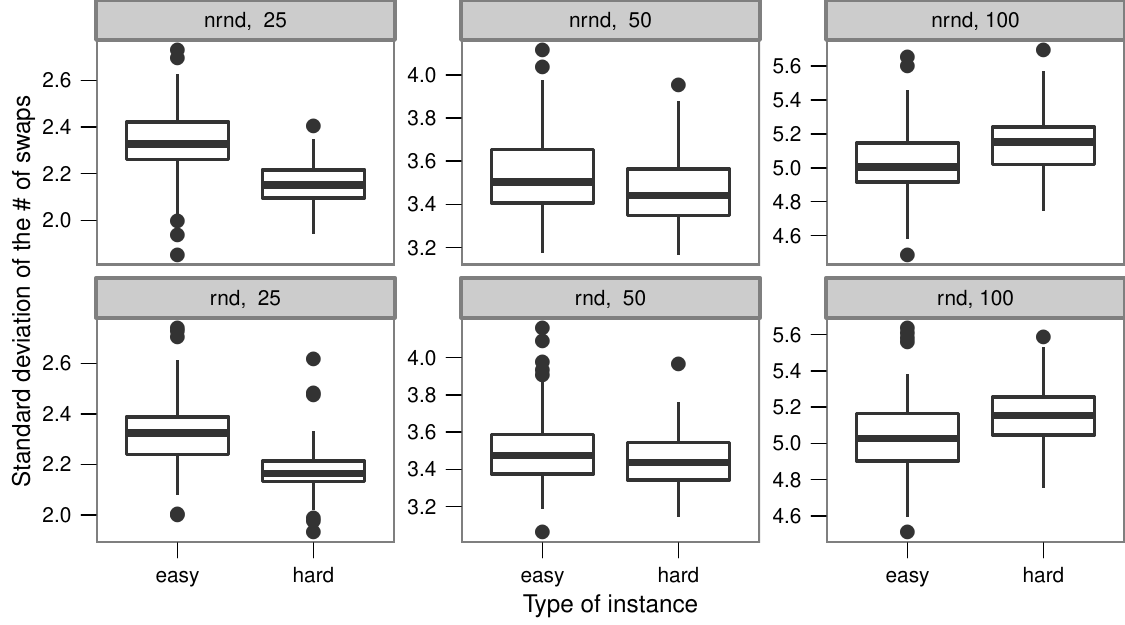}
  \caption{Mean number of swaps (top) executed by $2$-opt and standard
    deviations (bottom) for different instance sizes and rounding
    schemes.}\label{fig:swaps}
\end{figure}

Additionally, by analyzing the mean and standard deviation of the
number of swaps executed by $2$-opt (see Fig.~\ref{fig:swaps}) the
choice of choosing the approximation quality instead of the number of
swaps as suggested in \cite{SMHL10} as a meaningful performance
indicator for $2$-opt can be justified. It becomes obvious that there
is no positive correlation between the number swaps and problem
hardness measured by approximation quality. On the contrary, the
opposite trend can be observed. However, it is questionable if this
significant difference is really a relevant difference as the absolute
deviations in the number of swaps are very small. Therefore, at least
for $2$-opt, the number of swaps is not an adequate indicator for
algorithm performance.

\subsection{Classification of Instance Hardness}
\label{sec:class}

\begin{figure}[tb]
  \centering
  \includegraphics{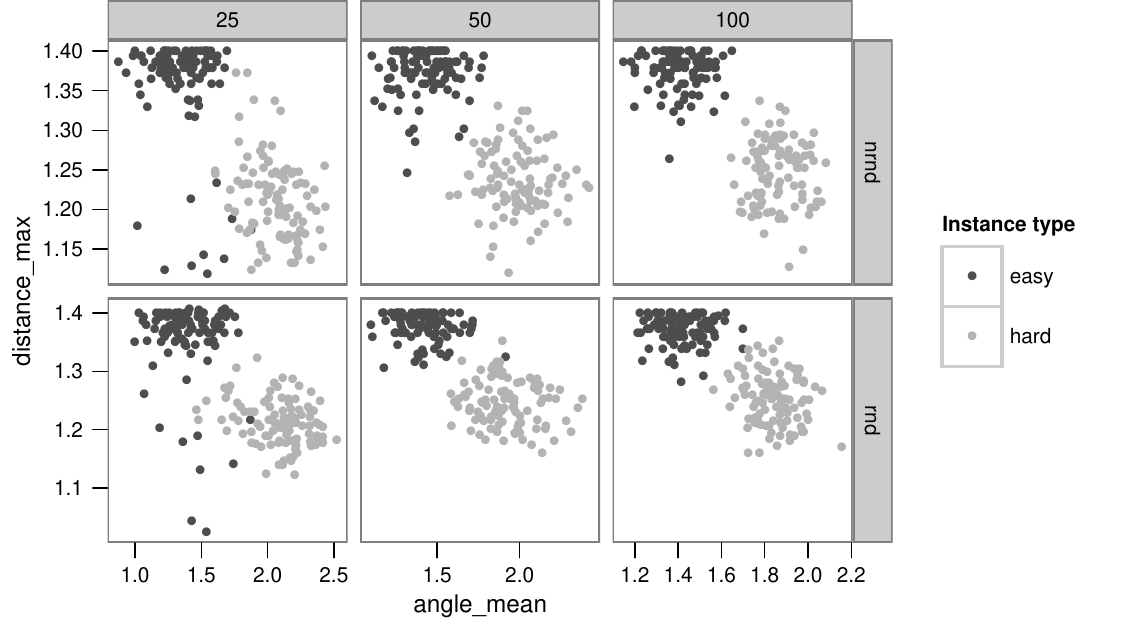}
  \includegraphics{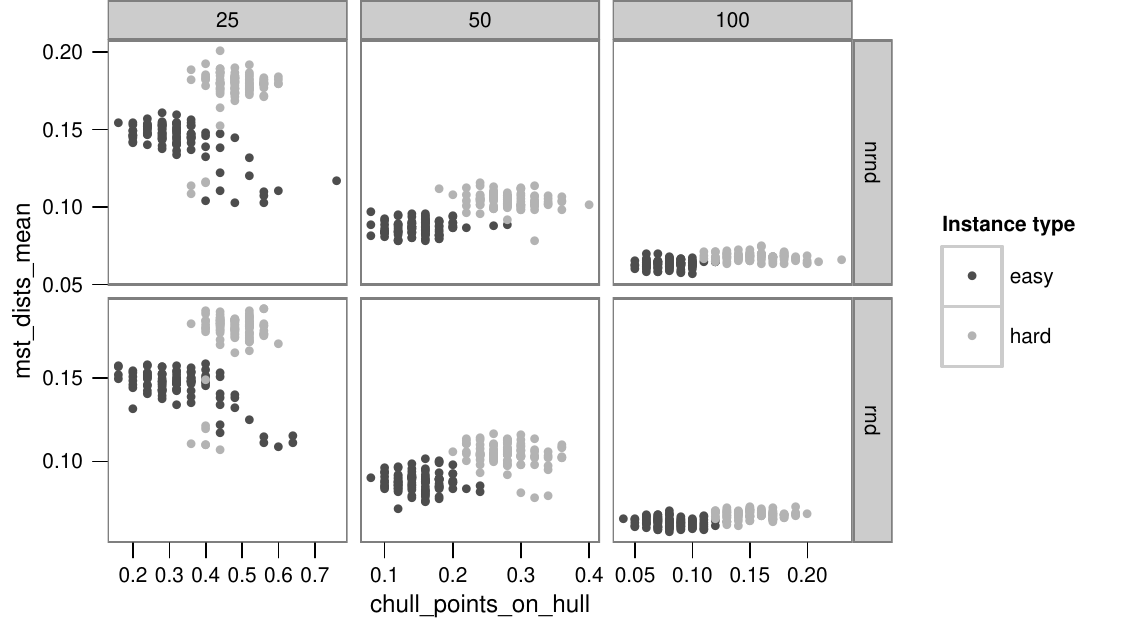}
  \caption{Scatt1erplot of exemplary feature combinations which allow
    an accurate separation of easy and hard instances. }
  \label{fig:class2d}
\end{figure}

A decision tree \cite{Breiman1984} is used to differentiate between
the two instance classes. Independent from the instance sizes and
rounding schemes an almost perfect classification of instances into
the two classes based on only two features is
possible. Fig.~\ref{fig:class2d} visualizes the values of two
exemplary feature combinations which can be used for this purpose. It
becomes obvious that the classification task is almost trivial as the
instance classes could be separated in a quite satisfactory manner
with one feature already.

\begin{figure}
  \centering
  \includegraphics{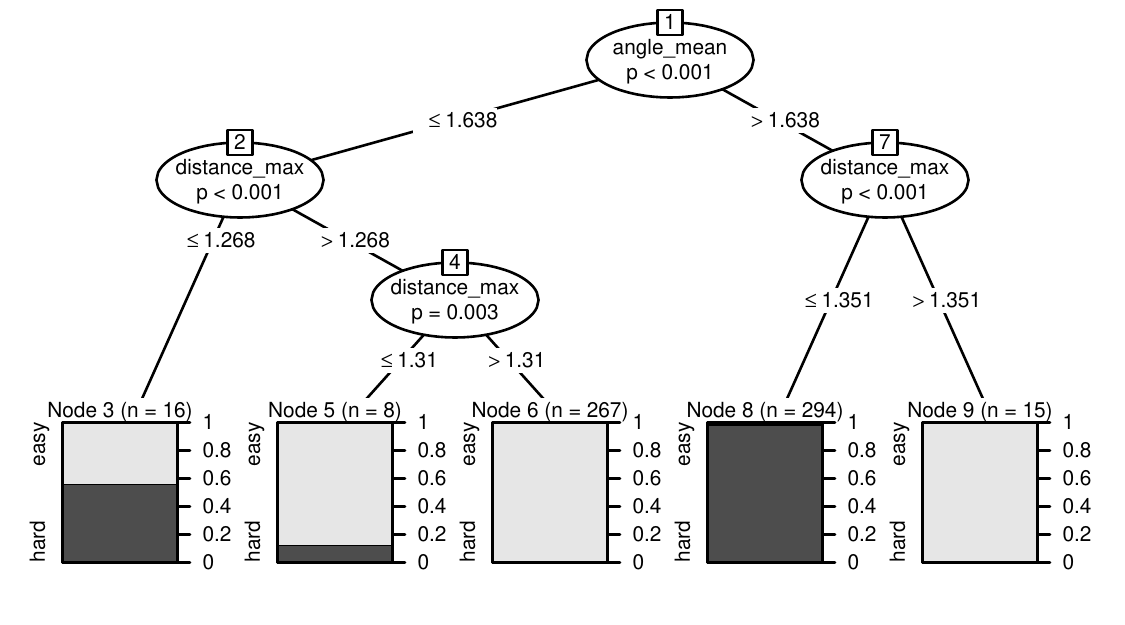}
  \caption{Classification rules for $rnd$ for the first feature
    combination given in Fig.~\ref{fig:class2d}. Mean classification
    accuracy equals $0.968$.}
  \label{fig:classrules1}
\end{figure}

\begin{figure}
  \centering
  \includegraphics[angle=-90,width=\textwidth]{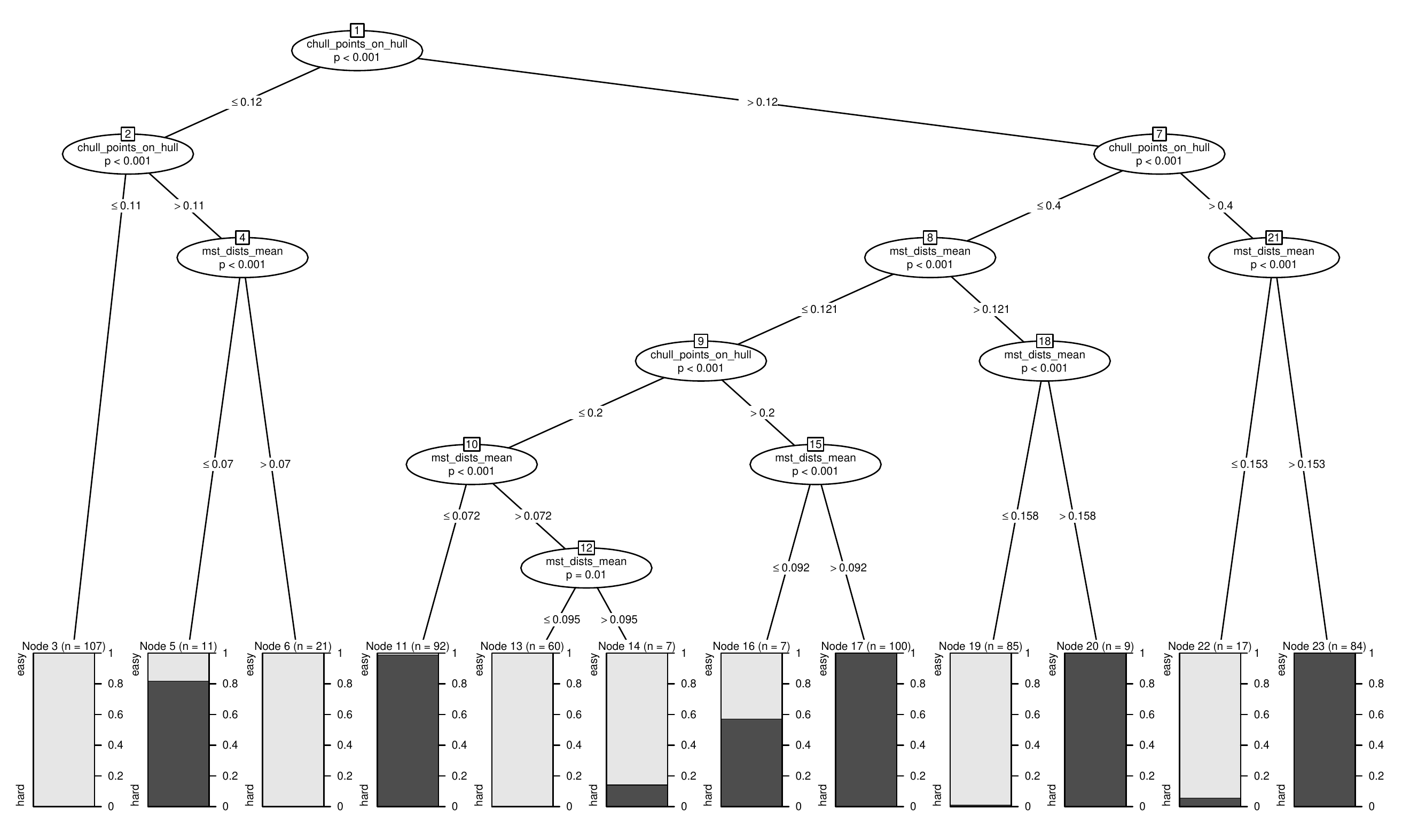}
  \caption{Classification rules for $rnd$ for the second feature
    combination given in Fig.~\ref{fig:class2d}. The cross-validated
    accuracy of this rule is $0.975$.}
  \label{fig:classrules2}
\end{figure}

The corresponding classification rules are presented in
Figs.~\ref{fig:classrules1} and \ref{fig:classrules2} for $rnd$ together with the ten-fold cross-validated classification accuracies.

The first rule is perfectly in line with the exploratory observations
of Section~\ref{sec:charInst}. The mean angles between the cities on
the optimal tour were found to be significantly higher for the hard
instances than the for the easy ones (see Fig.~\ref{fig:anglestour})
which is reflected by the corresponding classification rule. Secondly, the
easy instances exhibit a more uniform distribution of the tour length
legs on the optimal tour (see Fig.~\ref{fig:sdtour}). This observation
coincides with the characteristics of the feature dist\_max. The
higher the maximum distance between two cities the lower is the
probability of a low standard deviation of the tour length legs.

The second rule comprises two of the new features introduced in this
paper. A lower fraction of points on the convex hull of all points
together with smaller mean distance of the minimum spanning tree
indicates a low instance hardness. In general the rules are more
accurate for the higher instances sizes than for instance size 25.

Summarizing, an accurate feature-based separation of the easy and hard
instances can be successfully achieved, even with various combinations
of two features. Arising from this, we will investigate in the next
sections how instances of moderate difficulty in between the evolved
easy and hard instances can be generated and if an explicit prediction
of the expected approximation quality of $2$-opt on all instances is
possible based on the available features. This prediction problem in
our view is much more interesting as well as challenging than the
classification task analyzed within this section and will thus be
investigated in more detail.

\subsection{Morphing Hard into Easy Instances}

We are now in the position to separate easy and hard instances with
the classification rules presented in Section~\ref{sec:class}.  In this
section, instances in between, i.e. of moderate difficulty, are
considered as well. Starting from the result in \cite{EnglertRV07}
that a hard TSP instance can be transformed into an easy one by slight
variation of the node locations, we studied the ``transition'' of hard
to easy instances by morphing a single hard instance into an easy
instance by a convex combination of the points of both instances,
which generates an instance in between the original ones
(Alg.~\ref{alg:morph}).

\begin{algorithm}[h]
  \caption{Morphing}
  \label{alg:morph}
  \begin{algorithmic}
    \Function{morph}{$hardInstance,easyInstance,\alpha,cells,rnd$}
    \State $easyInstance \gets pointMatching(hardInstance,easyInstance)$
    \State $instMorph \gets \alpha\cdot hardInstance + (1-\alpha)\cdot easyInstance$
    \State $instMorph \gets \Call{rescale}{instMorph}$
    \State $instMorph \gets \Call{round}{instMorph,cells}$
    \If{$rnd$}
    \State $instMorph \gets \Call{normalMutation}{instMorph}$
    \State $instMorph \gets \Call{CutToBoundary}{instMorph}$
    \EndIf
    \State \Return $instMorph$
    \EndFunction
  \end{algorithmic}
\end{algorithm}

The point matching between the input instances is improved
w.r.t. \cite{mersmann2012} in that a greedy approach is used which
successively selects the point pair for which the pairwise Euclidean
distance is minimal. Depending on the type of rounding scheme used in
the EA that generated the instances, a normal mutation step and successive cutting to the boundary might be
required after rescaling and rounding to ensure that the newly
constructed instance is of the desired instance type.

\begin{figure}
  \centering
  \includegraphics[width=0.54\textwidth]{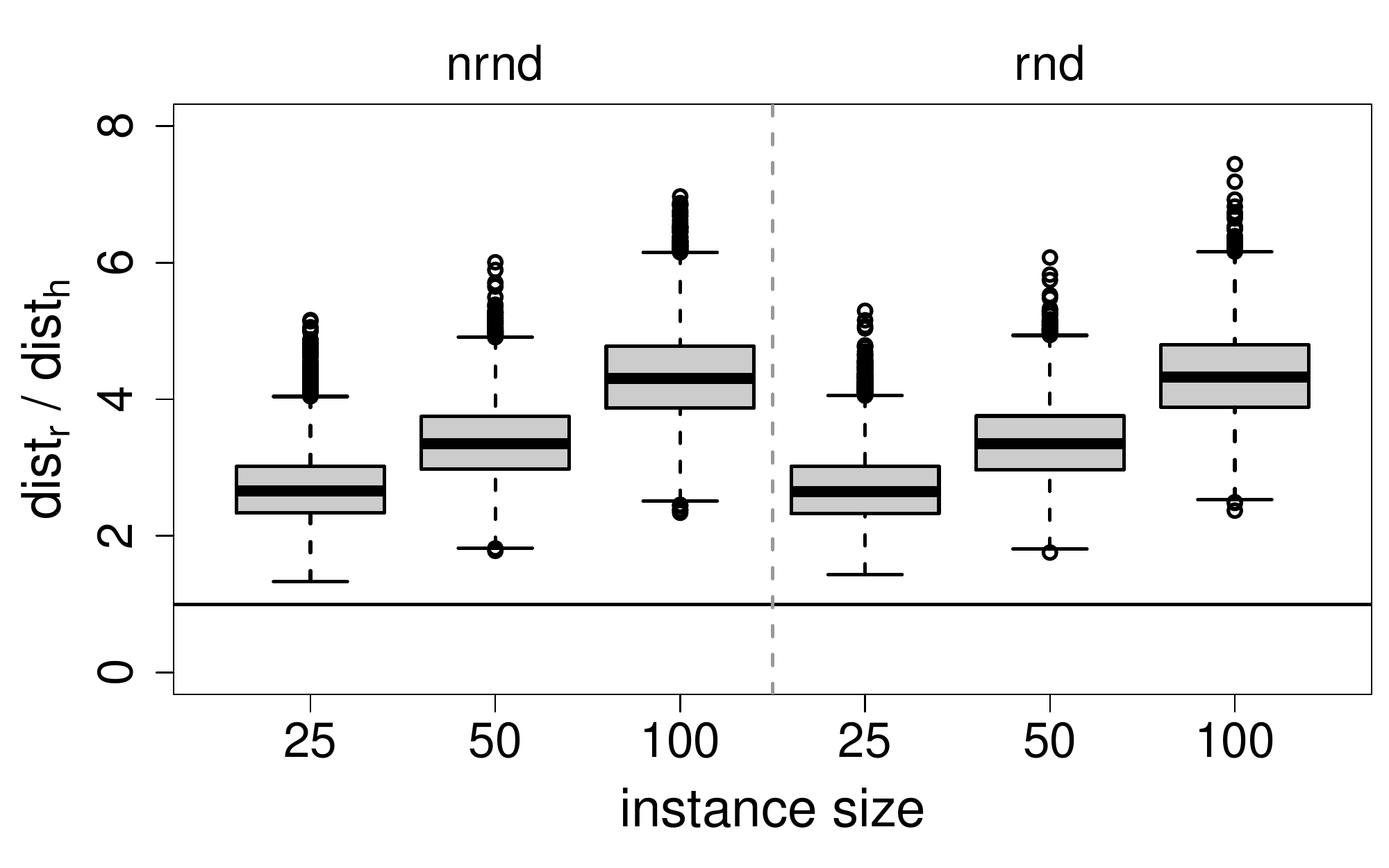}
  \caption{Effect of heuristic vs. random point matching
    strategy. Boxplots of the sums of all interpoint distances of the
    random approach (dist$_r$) relative to the heuristic ones
    (dist$_h$) are given for the two different rounding concepts and
    varying instance sizes.}
  \label{fig:effectmatch}
\end{figure}

Fig.~\ref{fig:effectmatch} shows the positive effect of the greedy
heuristic point matching strategy in contrast to a random point
matching as utilized in \cite{mersmann2012}. A simulation was
conducted in the following way: Two random instances are generated in
the $[0,1]^2$-plane, rescaled, rounded and (possibly) normally mutated
afterwards to reflect the EA rounding scheme which applies rounding
before normal mutation ($rnd$). Afterwards, the sum of interpoint
distance after random point matching (dist$_r$) and greedy heuristic
point matching ($dist_h$) are calculated and divided by each other
(dist$_r$ / dist$_h$). Results are shown for instance sizes
$\{25,50,100\}$ and both rounding schemes. It becomes obvious that the
interpoint distances resulting from the greedy approach are much
smaller than the respective ones of the random strategy. The latter
distance sums are on average roughly twice as high for the instance
size 25 and increases linearly to a factor of four for instance size
100. The effects are visually identical for both rounding concepts.

\begin{figure}
  \centering
  \includegraphics[width=0.24\textwidth]{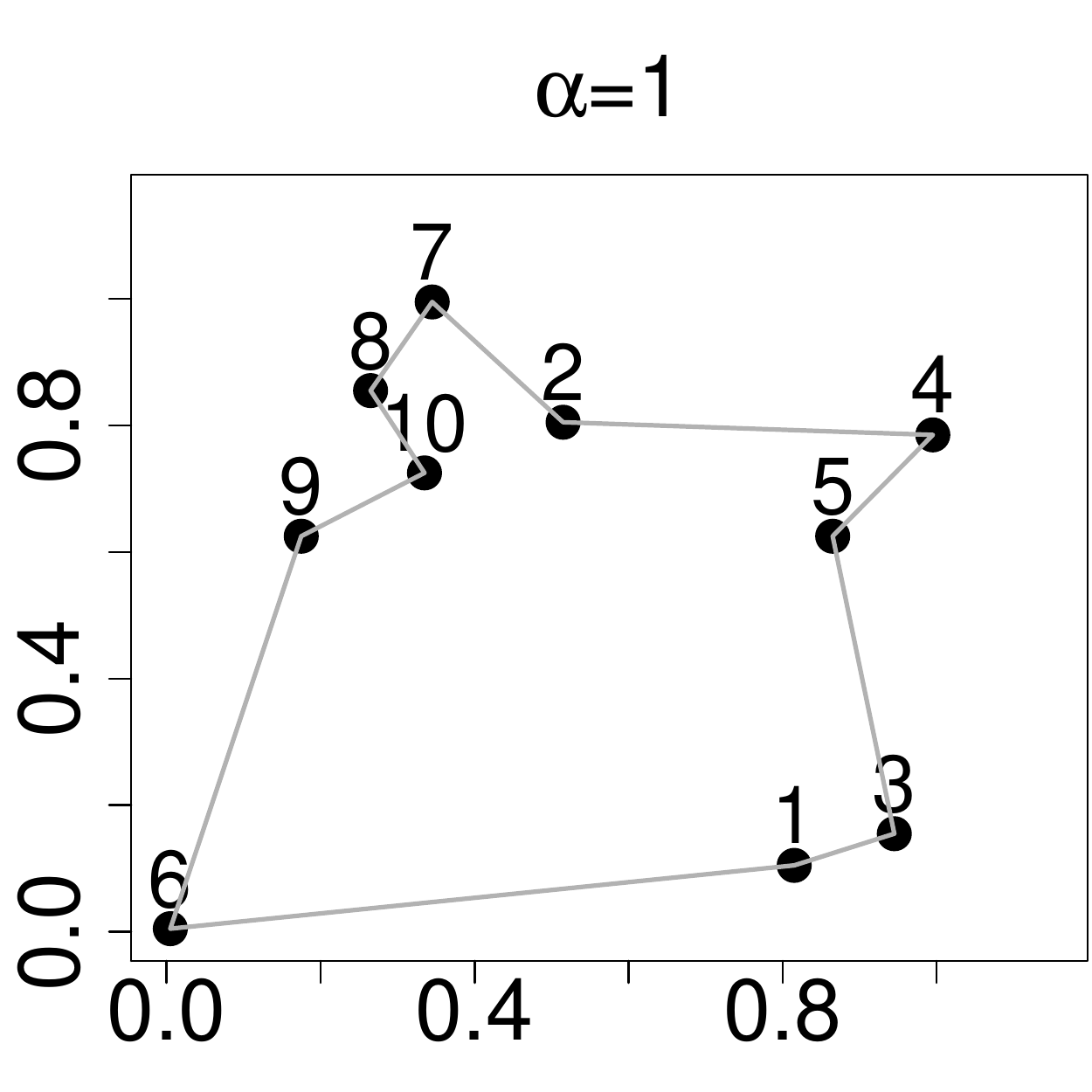}
  \includegraphics[width=0.24\textwidth]{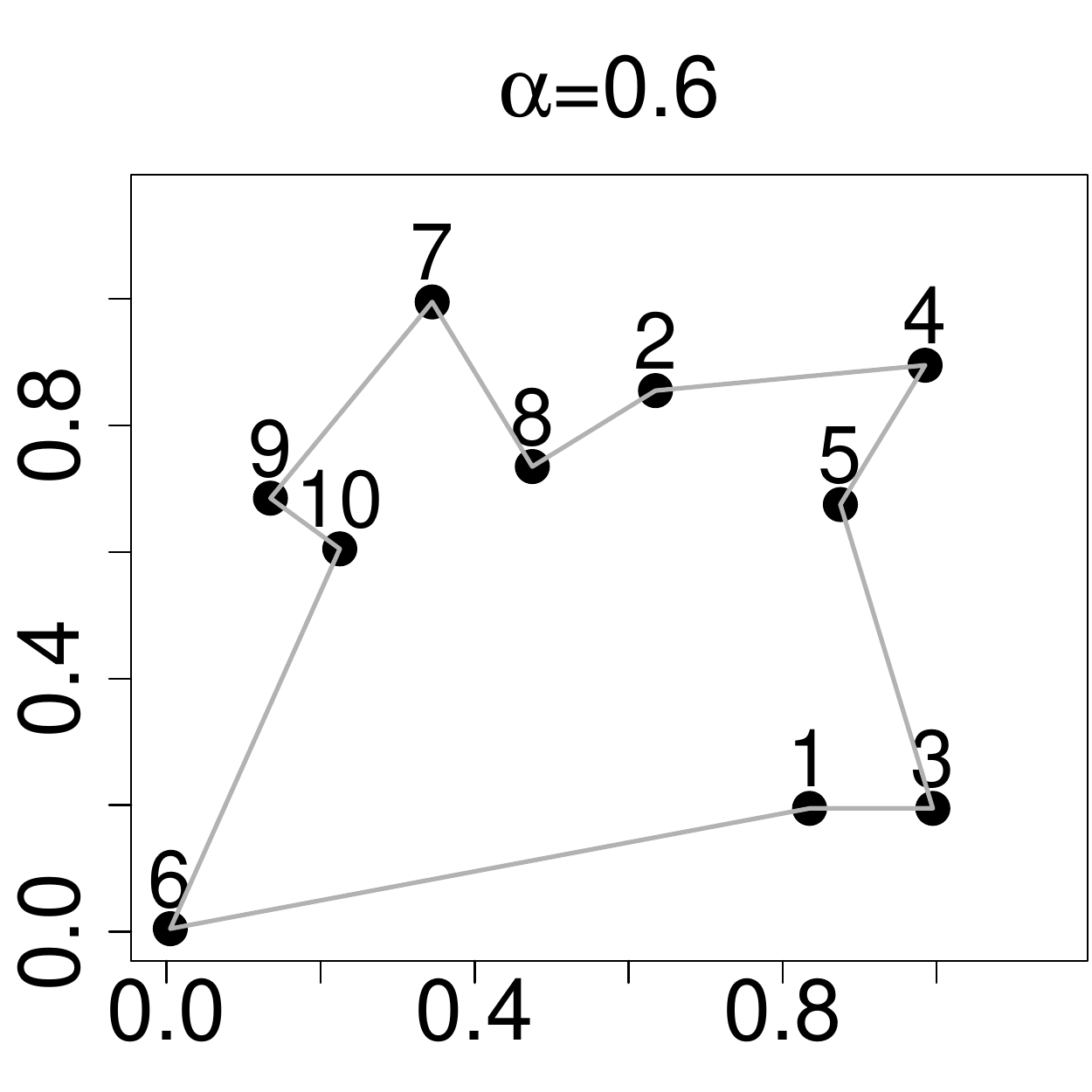}
  \includegraphics[width=0.24\textwidth]{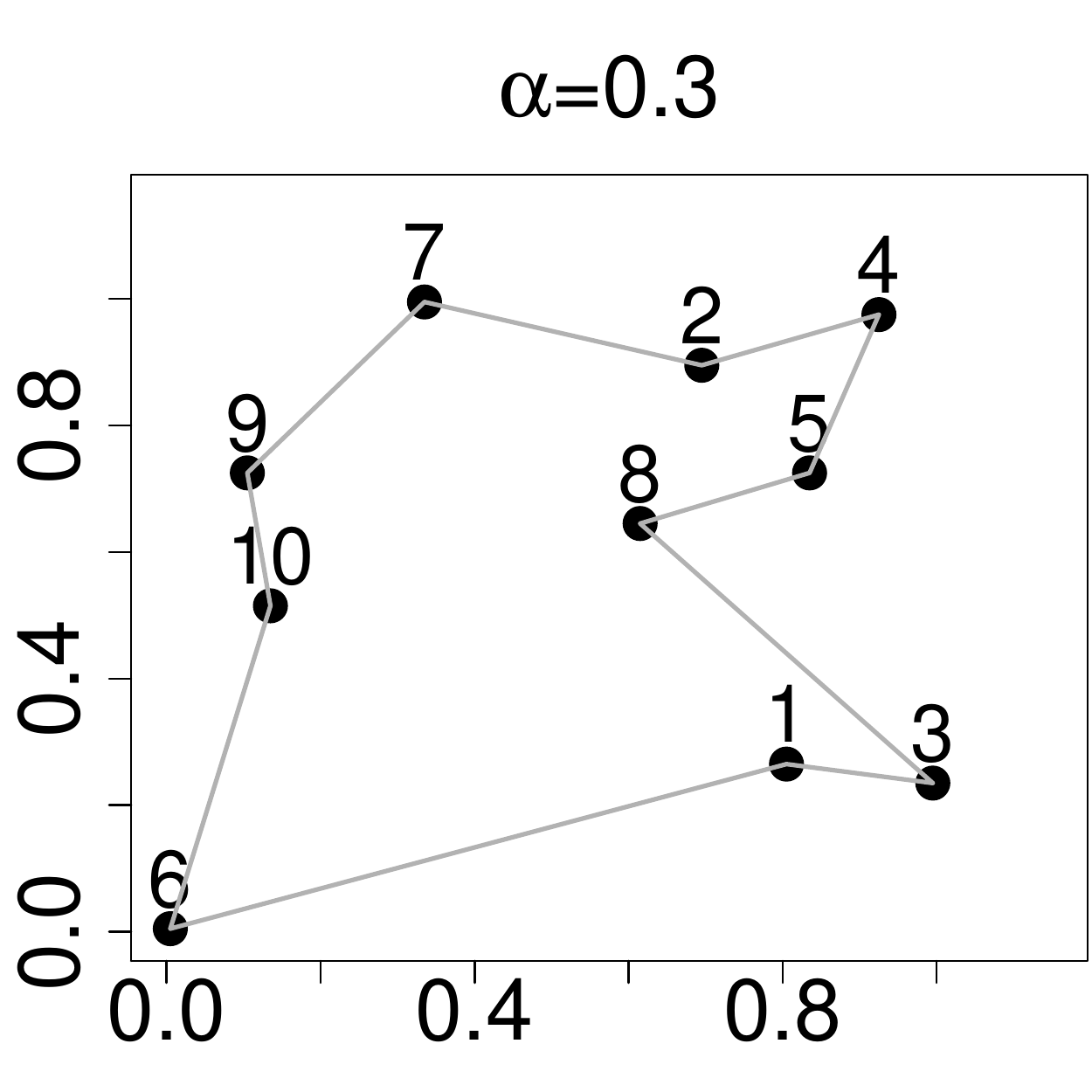}
  \includegraphics[width=0.24\textwidth]{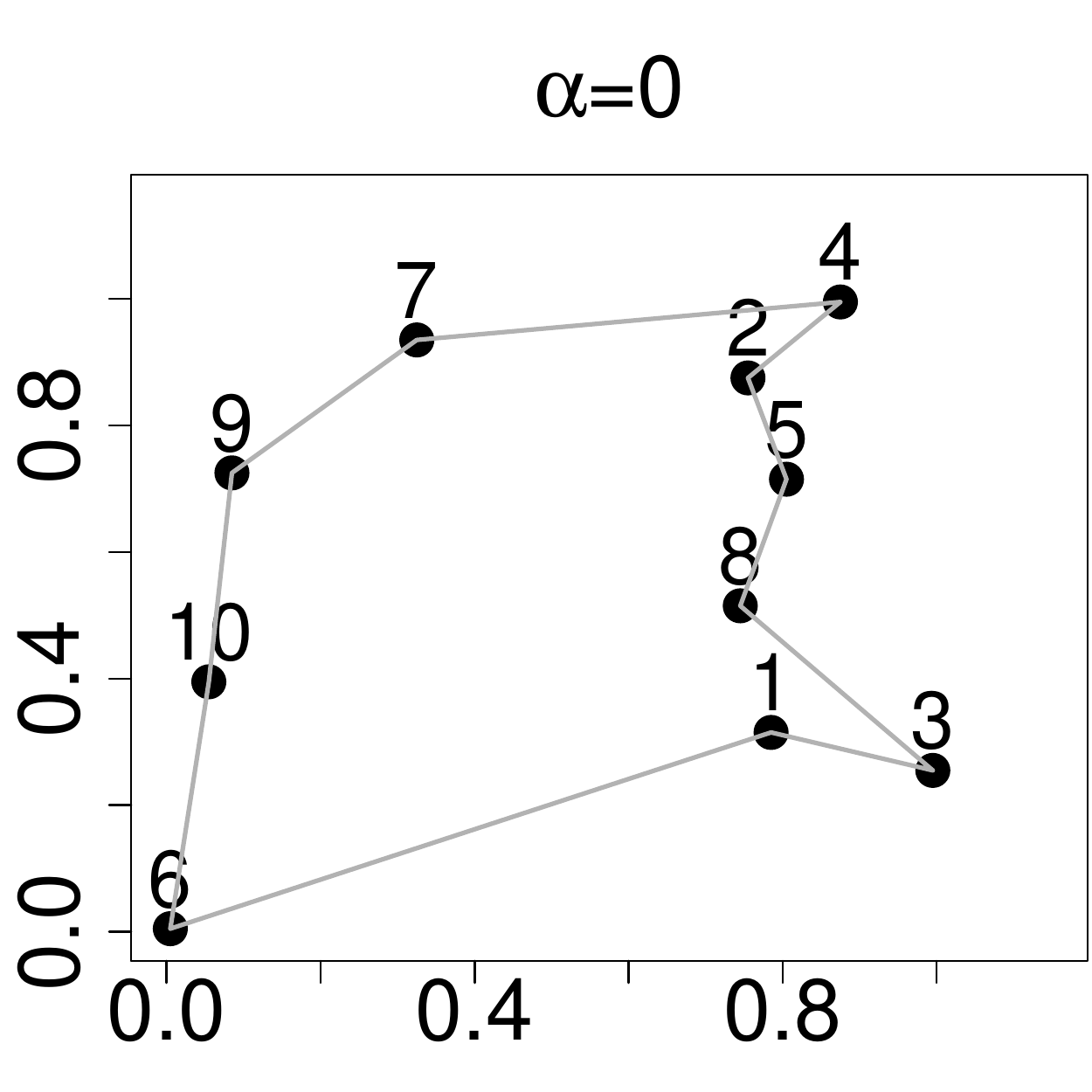}\\
  \includegraphics[width=0.24\textwidth]{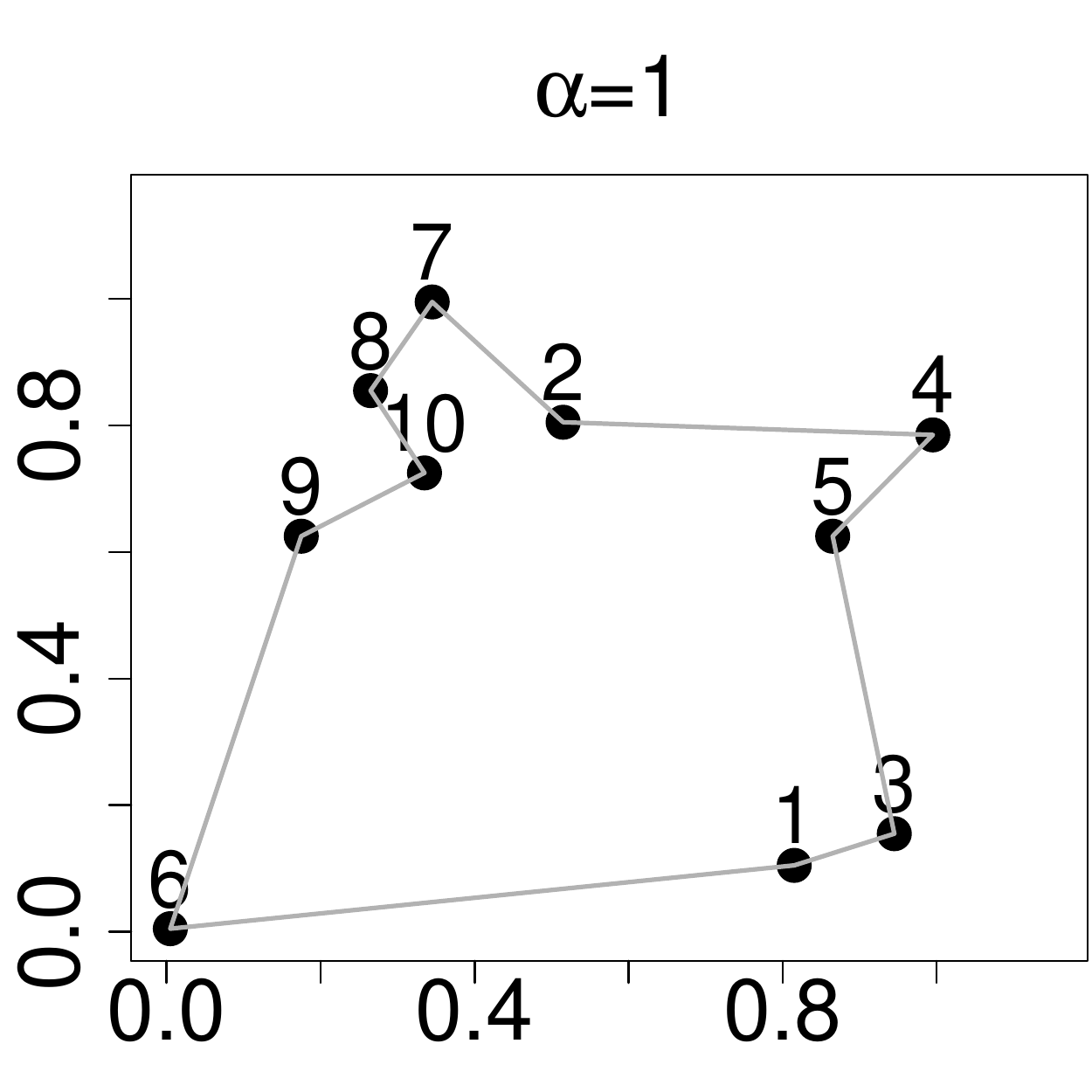}
  \includegraphics[width=0.24\textwidth]{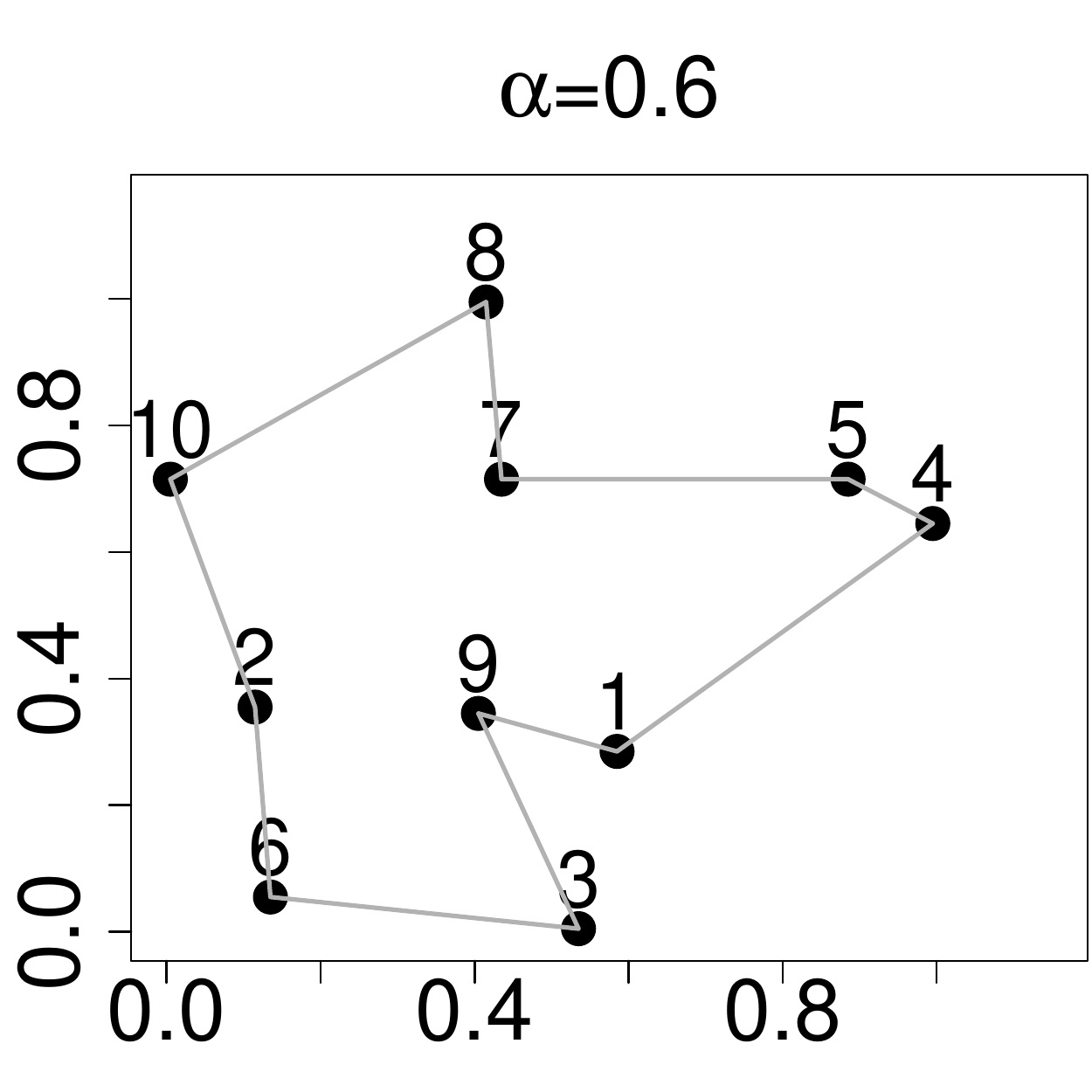}
  \includegraphics[width=0.24\textwidth]{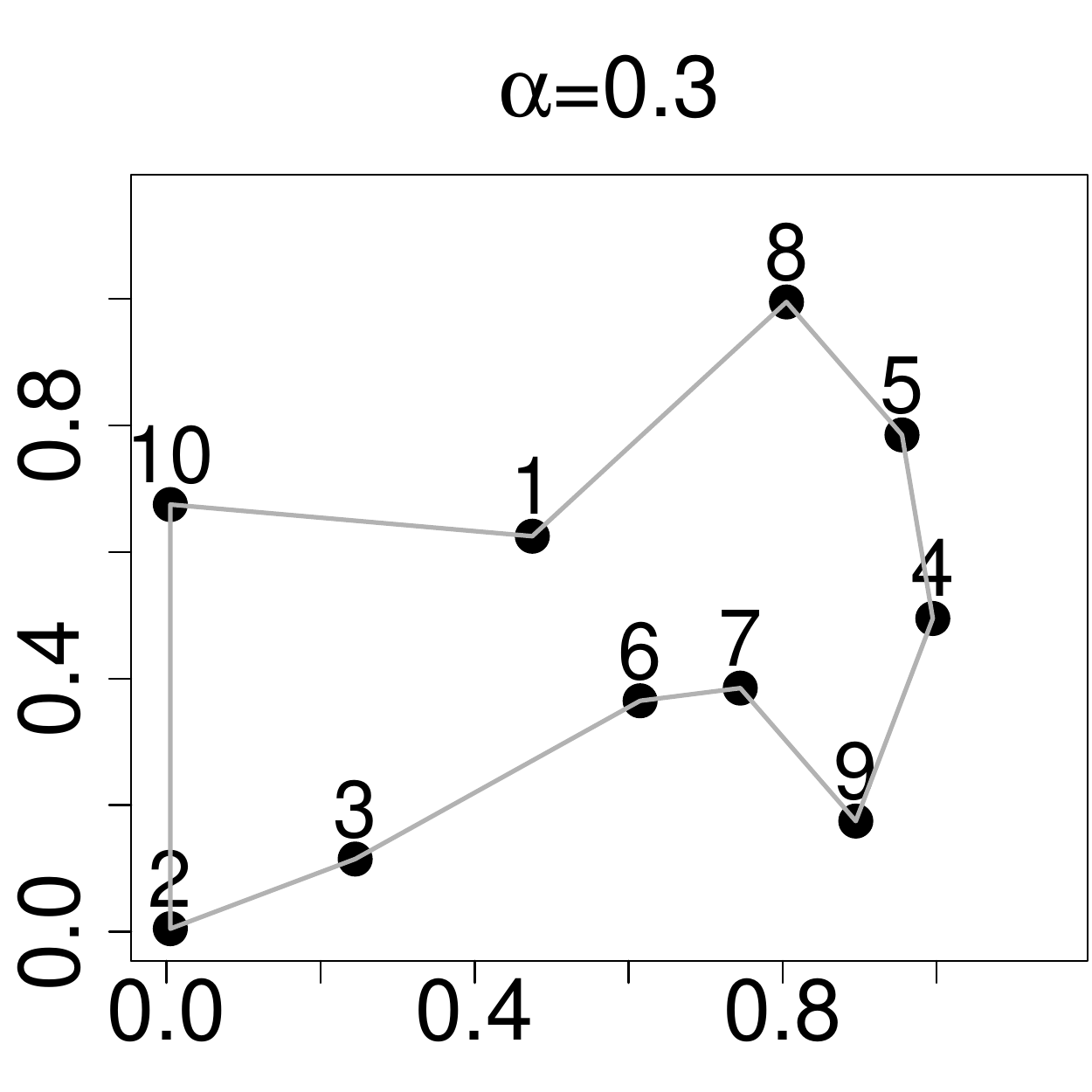}
  \includegraphics[width=0.24\textwidth]{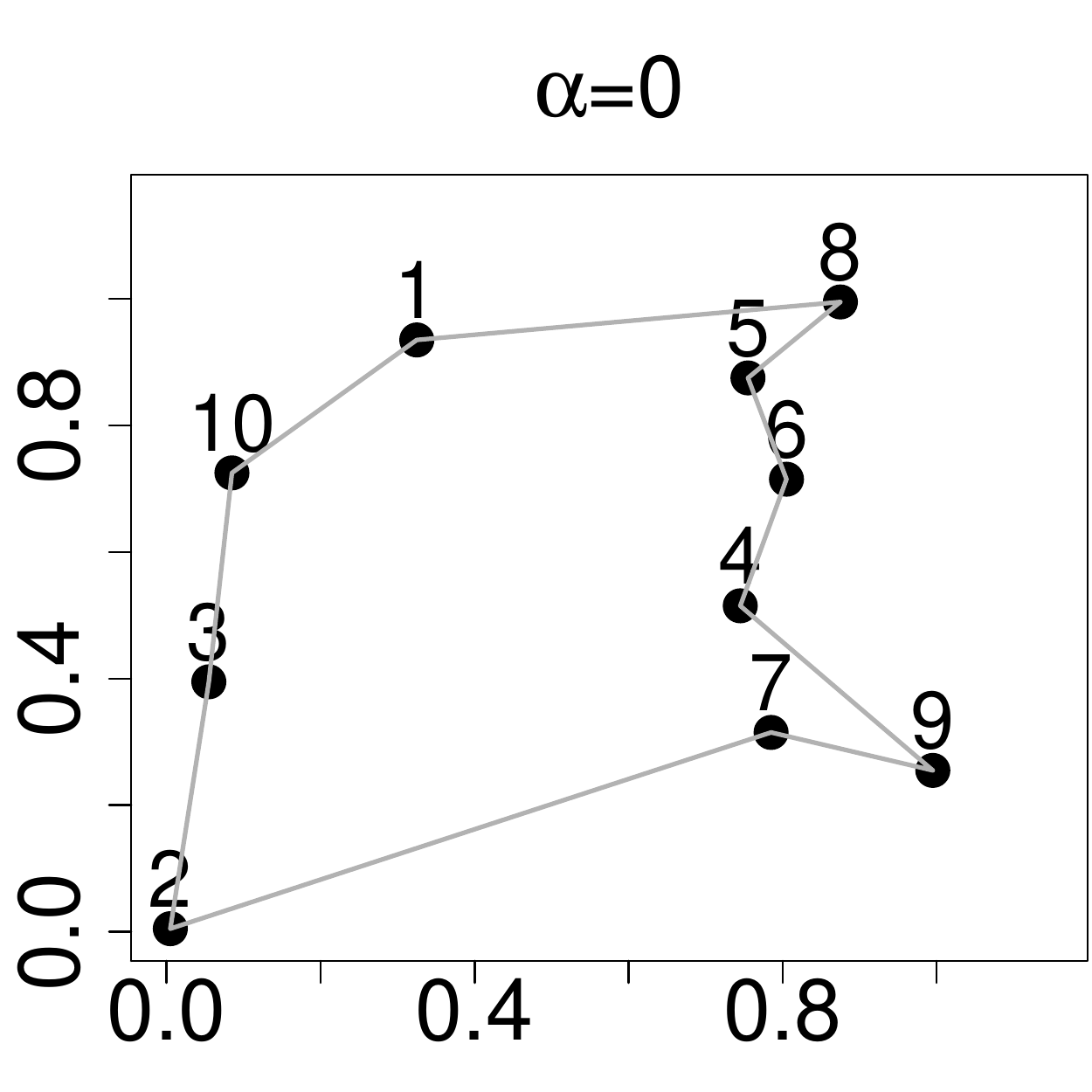}\\
  \caption{Example: Morphing of one instance into a different instance
    for different $\alpha$-levels of the convex combination ($nrnd$)
    with heuristic greedy (above) and random point matching
    (below). Optimal tours are visualized in grey.}
  \label{fig:morphsketch}
\end{figure}

Morphing examples are shown in Figure~\ref{fig:morphsketch}. Based on
the initial instances ($\alpha=1$ and $\alpha=0$), instances emerge
from each other with decreasing $\alpha$. Clearly, the advantageous
effect of the improved point matching becomes visible as the
transitions of the morphed instances are much smoother than in the
former case of random point matching. Furthermore, in case of random
point matching instances tend to concentrate on the center part of the
$[0,1]^2$ - plane.

The morphing strategy is applied to all possible combinations of single
hard and easy instances of the two evolved instance sets using 6
levels of $\alpha$, i.e. $\alpha \in \{0,0.2,0.4,...,1\}$. Each
generated instance is characterized by the levels of the features
discussed in Section~\ref{subsec:feat}. Thus, the changes of the
feature levels with increasing $\alpha$ can be studied which is of
interest as it should lead to an understanding of the influence of the
different features on the approximation quality.

\begin{figure}
  \centering
  \includegraphics{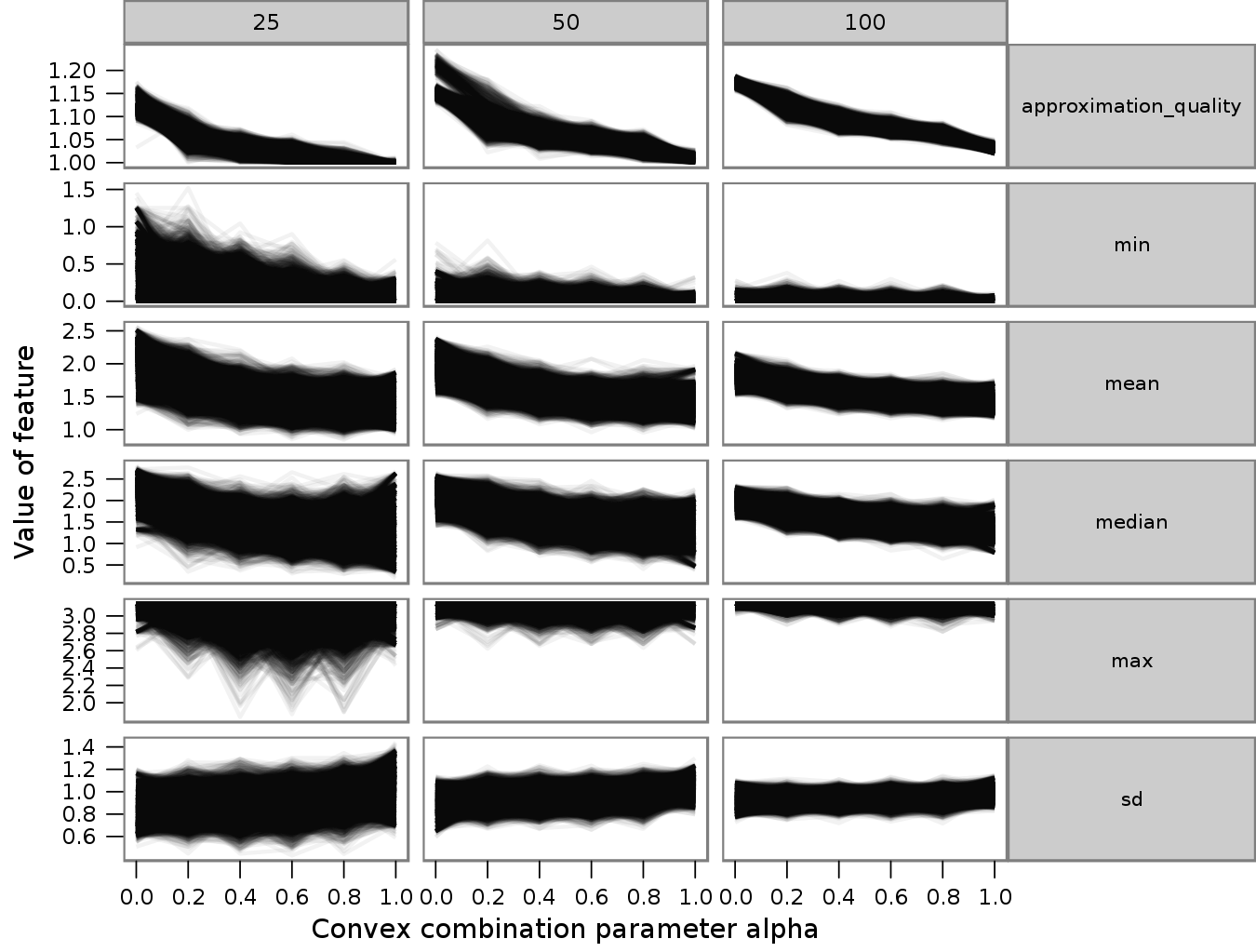} \\\vspace*{0.3cm}
  \includegraphics{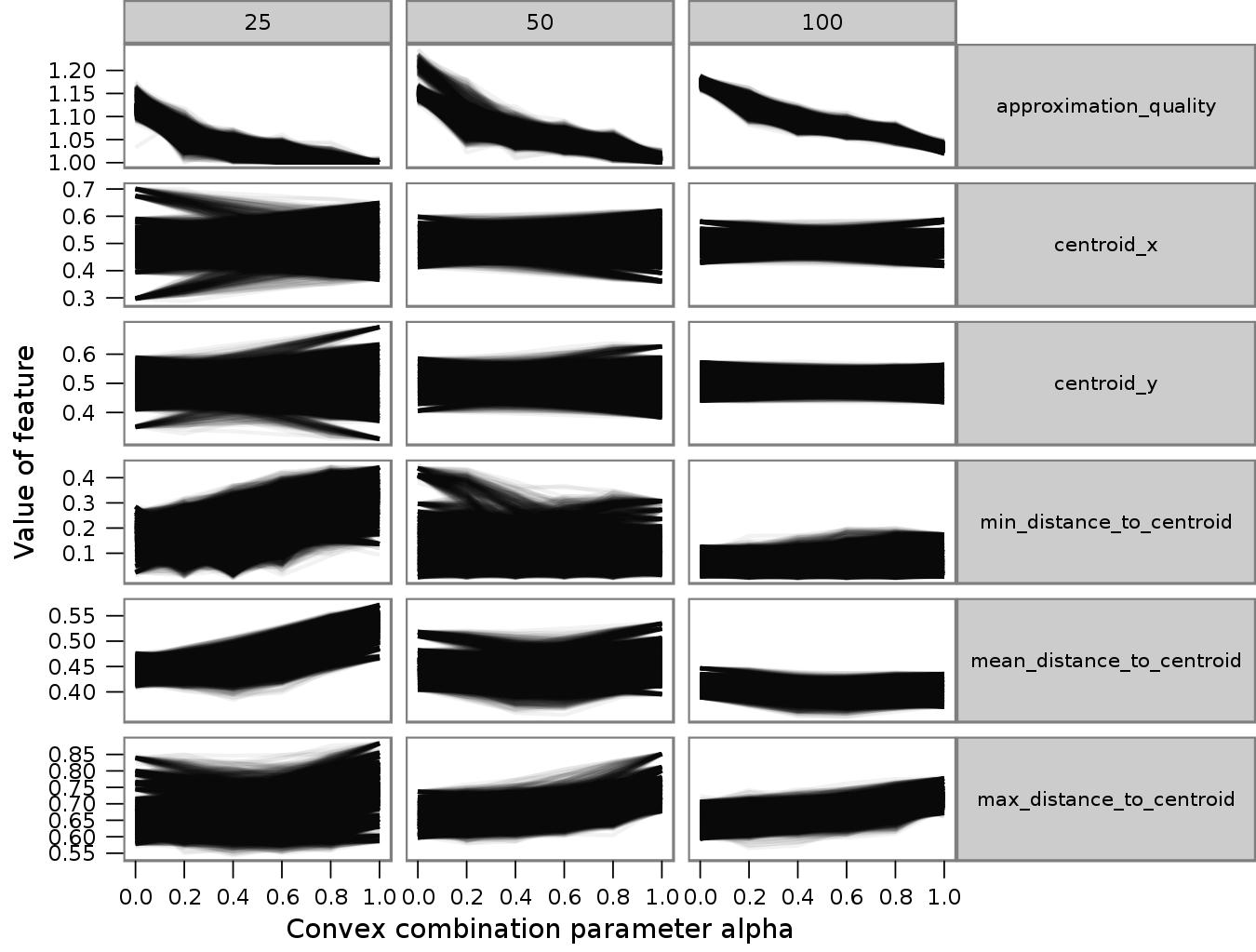}
  \caption{Angle (top) and Centroid Features (bottom): Approximation
    quality and feature values for different $\alpha$ levels of all
    conducted morphing experiments.}
  \label{fig:morph1}
\end{figure}

\begin{figure}
  \centering
  \includegraphics{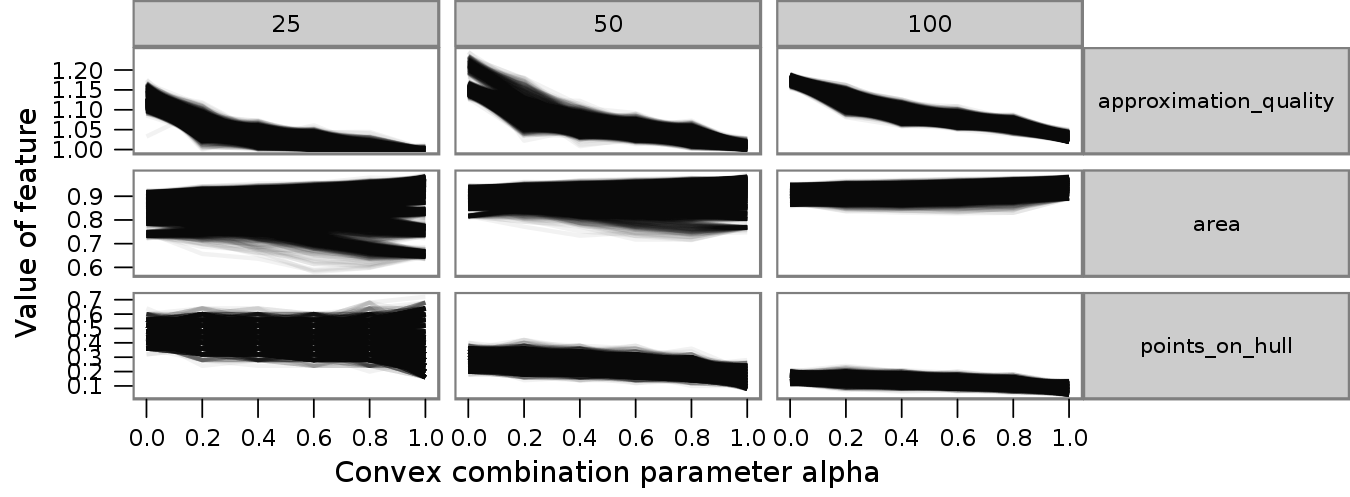}\vspace*{0.3cm}
  \includegraphics{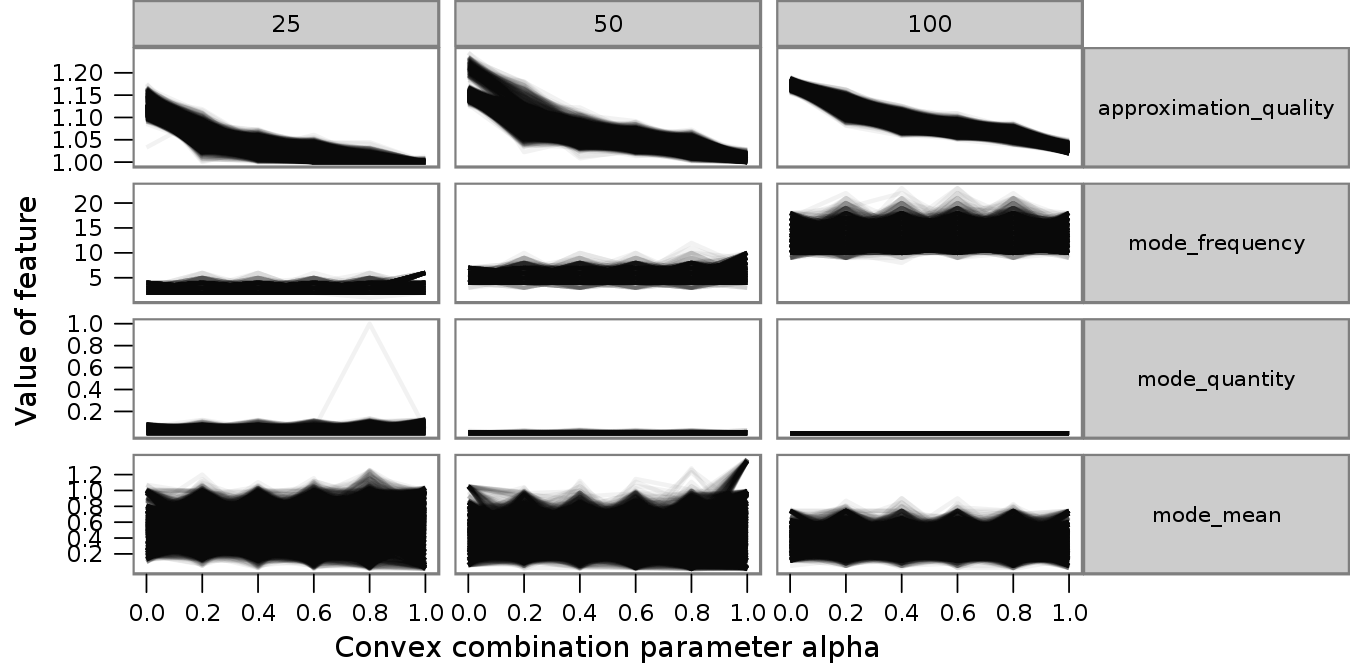}
  \caption{Convex Hull (top) and Mode (bottom) features: Approximation quality and feature values for different
    $\alpha$ levels of all conducted morphing experiments. }
  \label{fig:morph2}
\end{figure}

\begin{figure}
  \centering
  \includegraphics{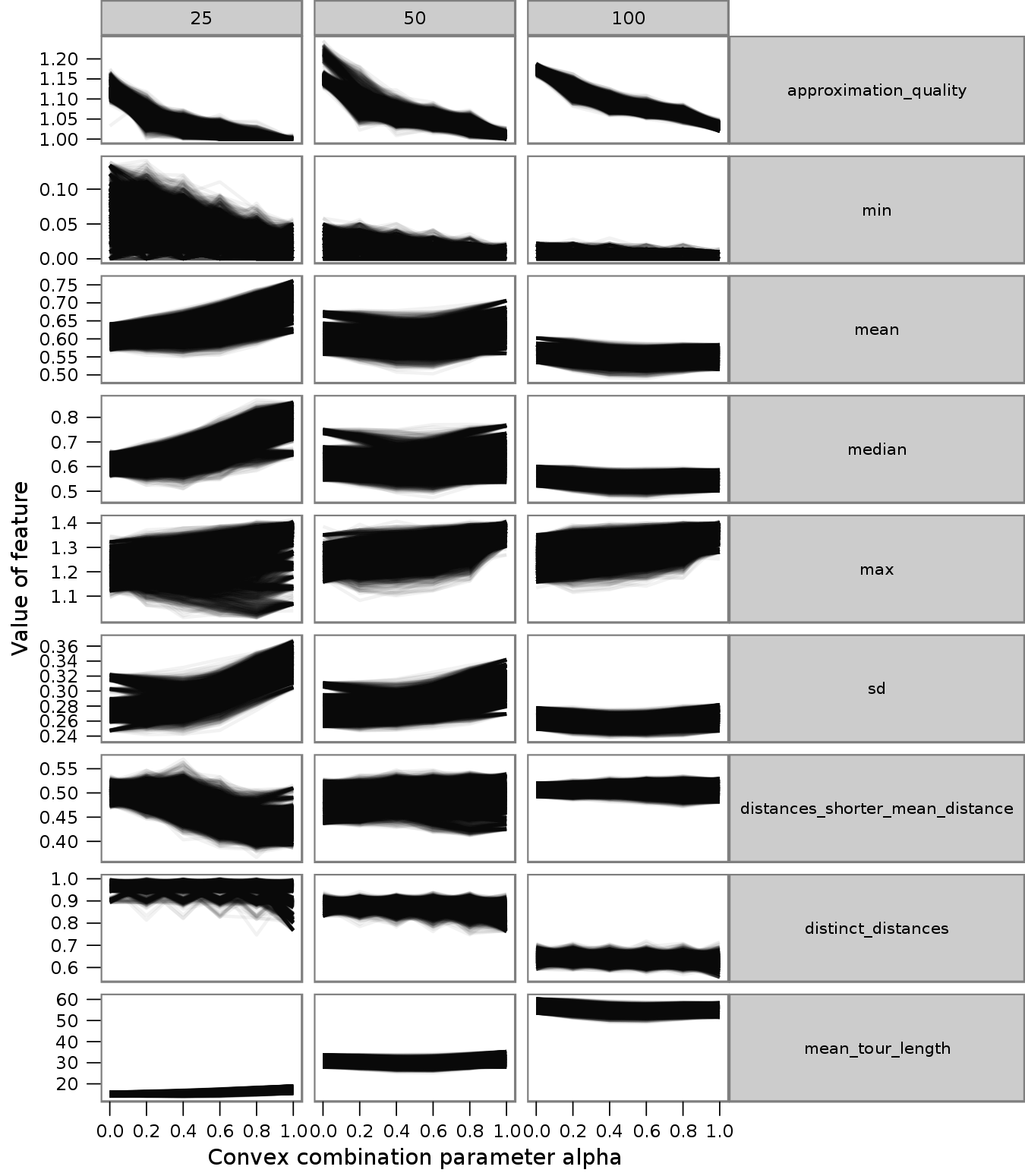}\\ \vspace*{0.3cm}
  \includegraphics{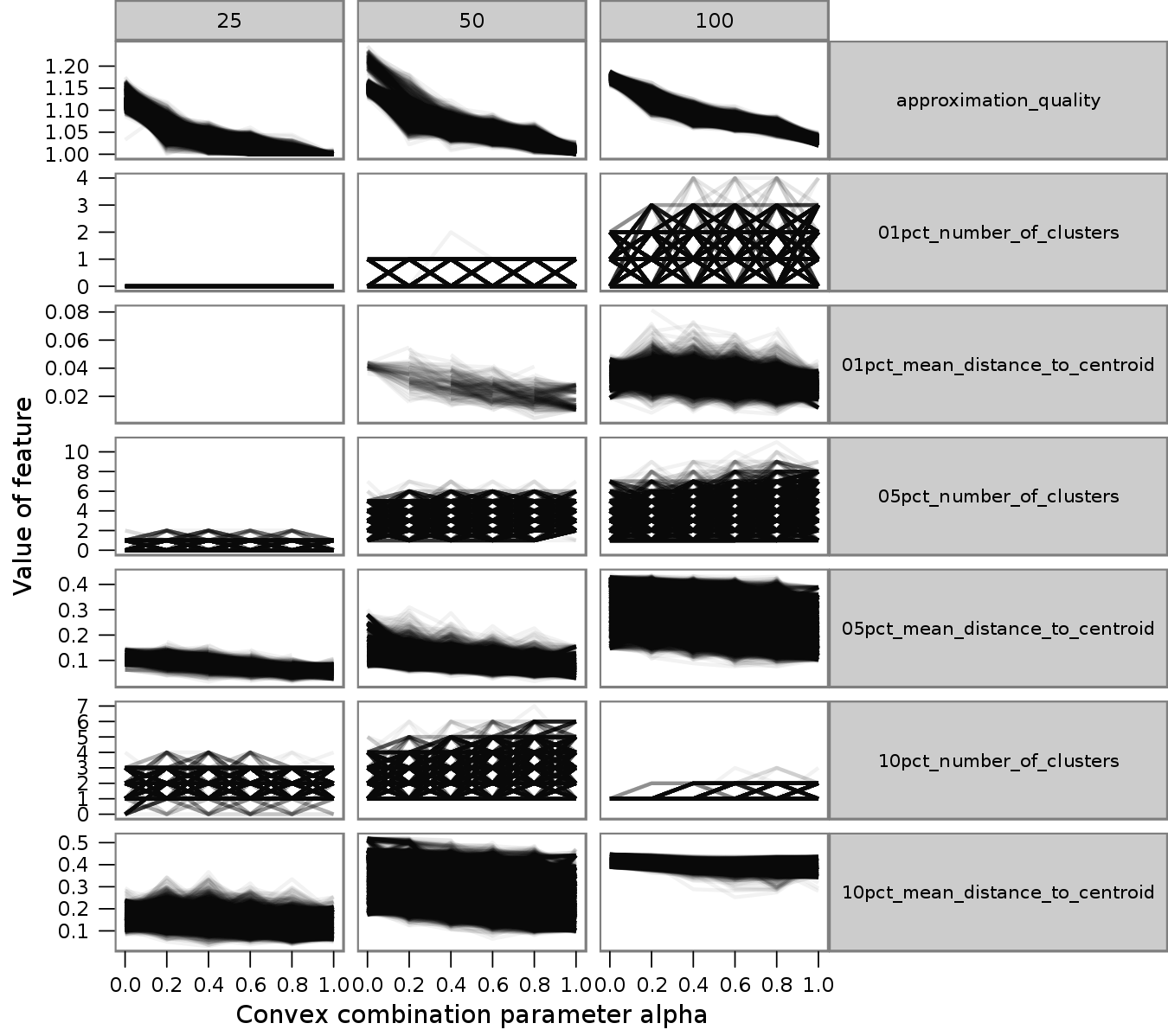}
  \caption{Distance (top) and Cluster (bottom) features: Approximation
    quality and feature values for different $\alpha$ levels of all
    conducted morphing experiments. The annotations ``01pct'',
    ``05pct'' and ``10pct'' identify different levels of the
    reachability distance as a parameter of GDBSCAN \cite{Sander98}
    used for clustering.}
  \label{fig:morph3}
\end{figure}

\begin{figure}
  \centering
  \includegraphics{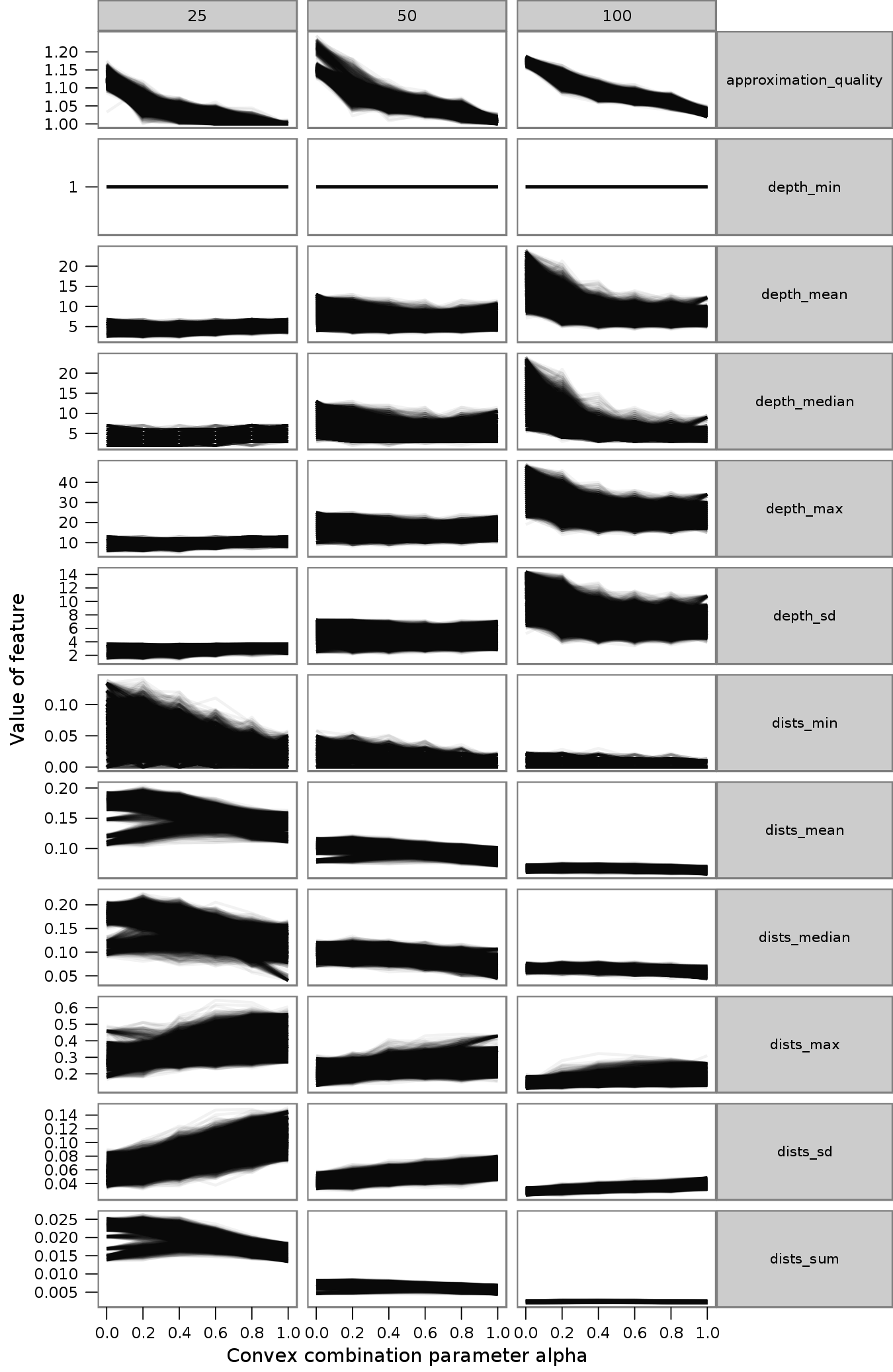}
  \caption{MST features: Approximation quality and feature values for
    different $\alpha$ levels of all conducted morphing experiments.}
  \label{fig:morph4}
\end{figure}

\begin{figure}
  \centering
  \includegraphics{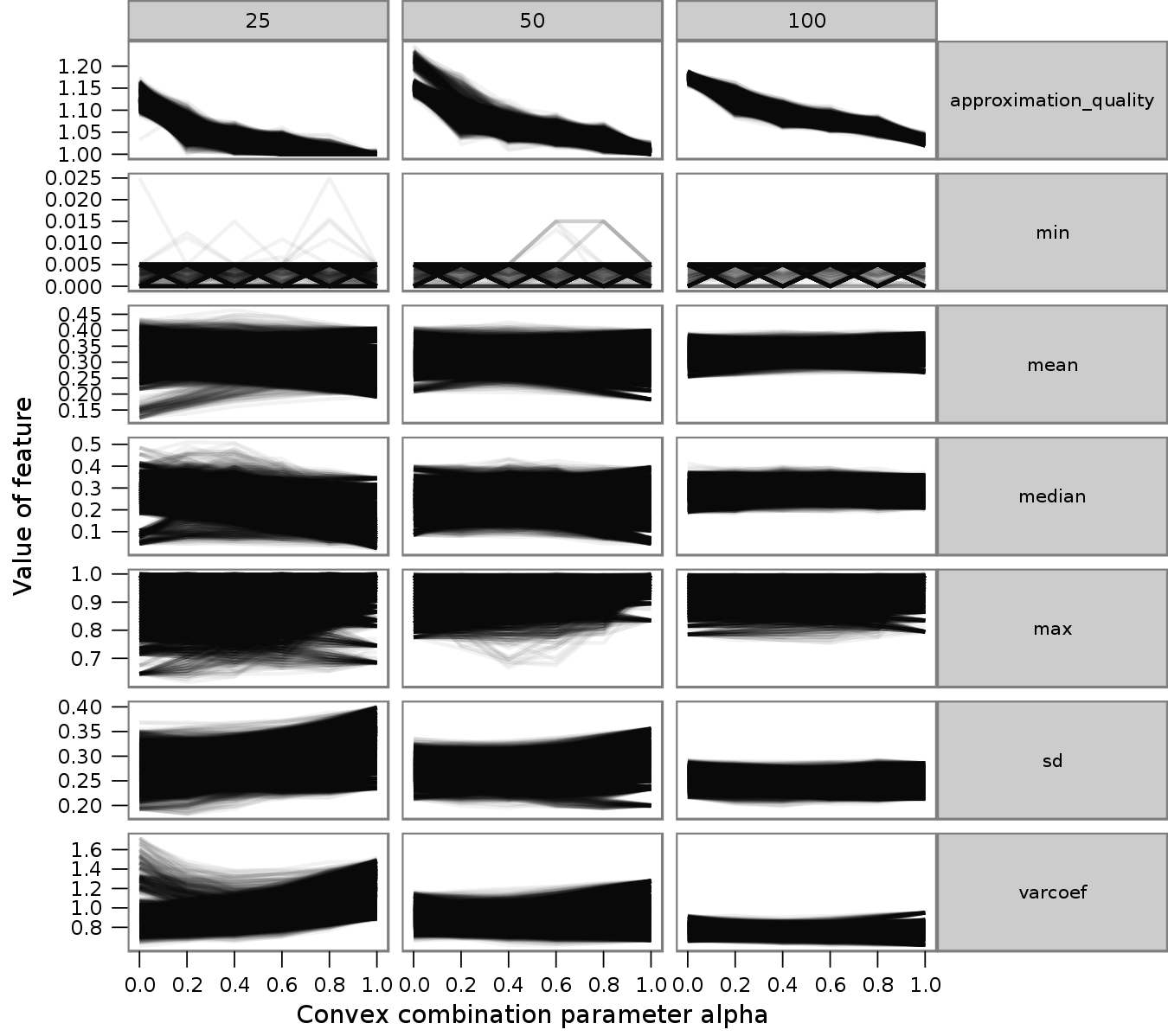}
  \caption{Nearest  neighbor distance features:  Approximation quality
    and feature values for  different $\alpha$ levels of all conducted
    morphing experiments.}
  \label{fig:morph5}
\end{figure}

Figures~\ref{fig:morph1} - \ref{fig:morph5} show the approximation
quality for the instances of all morphing sequences for the various
$\alpha$ levels in the top subfigure. Starting from a hard instance on
the left side of each individual plot ($\alpha=0$) the findings of
\cite{EnglertRV07} are confirmed. The approximation quality of $2$-opt
quickly increases with slight increases of $\alpha$. Additionally, the
feature levels of the generated instances are visualized arranged by
feature groups. We concentrate on the subset of data which is based on
rounding before the normal mutation step as the presented observations
coincide for both rounding schemes.

Obviously, many features do not show any systematic relationship with
the approximation quality for all considered instance sizes, e.g. most
features related to the centroid, the clustering as well as the modes
of the edge cost distribution. Interestingly, some features exhibit
different tendencies for the smallest and highest instance size,
e.g. the features reflecting the mean and minimum distance to the
centroid in Fig.~\ref{fig:morph1}. This is due to the different
structural shapes of small and large instance classes, more
specifically the almost circular structure of the easy instances for
an instance size of 25 (see Fig.~\ref{fig:extour}). The same reasoning
holds for the features associated with the mean and median distances
and the standard deviation of the distance matrix. Exceptional
behavior of the feature levels occurs for the features related to the
MST depth (mean, median, max and sd) for which a systematic nonlinear
decrease can be observed only for the instance size 100.  In
most cases the change of the feature in relation to $\alpha$ is
similar for all instance sizes, what does change is the variance of
the feature.

However, systematic nonlinear relationships with the approximation
quality can be detected for the mean and median distances on the MST
as well as the standard deviation (Fig.~\ref{fig:morph4}), the maximum
distance between the cities and the respective standard deviation (sd)
(Fig.~\ref{fig:morph3}) and the coefficient of variation of nearest
neighbor distances (Fig.~\ref{fig:morph5}). Additional promising
features in Fig.~\ref{fig:morph1} are the fraction of points on the
convex hull, the area of the convex hull, and the mean angle between
adjacent cities as well as the maximum distance to the centroid in
Fig.~\ref{fig:morph2}. Naturally, the features included in the two
exemplary classification rules above form a subset of the mentioned
relevant features.

\subsection{Feature-Based Prediction of TSP
  Problem Hardness}
\label{sec:MARS}

In order to get a more accurate picture of the relationship between
the approximation quality and the features a Multivariate Adaptive
Regression Splines (MARS) \cite{Friedman1991} model is constructed in
order to directly predict the expected approximation quality of
$2$-opt on a given instance based on the candidate features.

We used MARS with second degree interaction effects to model the
relationship between the approximation quality and the calculated
instance features. Other modeling approaches, such as $k$-nearest
neighbors and linear models, were also considered but some initial
experiments on a subset of the data showed that MARS provided
competitive results and scaled well to the full dataset. The final
model is shown in Table~\ref{tab:MARS}. We achieved a root mean
squared error (RMSE) of $0.016964$ for $rnd$ and $0.016502$ for
$nrnd$. This compares favorably to a simple model that always
predicts the mean (RMSE for $rnd$ equals $0.5115443$ and for $nrnd$
$0.0516154$) which we outperform by a factor of 3. In other words,
given the features of a TSP instance, we expect to predict, on
average, the approximation quality of a $2$-opt solution to within
$\pm 1.6\%$ of the true approximation ratio. In the following, we concentrate our analysis on $rnd$ as the results almost coincide.

\begin{table}
  \centering
\begin{tabular}{lr}
  \toprule
Spline & Coefficient \\ 
  \midrule
(Intercept) & 1.133 \\ 
  h(mst\_dists\_sd-0.0369208) & -0.716 \\ 
  h(0.0369208-mst\_dists\_sd) & 1.707 \\ 
  h(angle\_mean-1.45864) & 0.100 \\ 
  h(1.45864-angle\_mean) & 0.012 \\ 
  h(chull\_points\_on\_hull-0.28) & -0.012 \\ 
  h(0.28-chull\_points\_on\_hull) & -0.147 \\ 
  h(mst\_depth\_max-12) & 0.005 \\ 
  h(12-mst\_depth\_max) & 0.001 \\ 
  h(distance\_max-1.26151) & -0.199 \\ 
  h(1.26151-distance\_max) & 0.039 \\ 
  h(cluster\_10pct\_mean\_distance\_to\_centroid-0.496846) & -0.788 \\ 
  h(0.496846-cluster\_10pct\_mean\_distance\_to\_centroid) & -0.003 \\ 
  h(angle\_mean-1.45864)*h(distance\_mean-0.606019) & -0.795 \\ 
  h(angle\_mean-1.45864)*h(0.606019-distance\_mean) & -0.561 \\ 
  h(mst\_depth\_median-12) & -0.018 \\ 
  h(12-mst\_depth\_median) & 0.002 \\ 
  h(distance\_sd-0.307304) & -0.458 \\ 
  h(0.307304-distance\_sd) & 0.695 \\ 
  h(centroid\_max\_distance\_to\_centroid-0.72554)*h(1.26151-distance\_max) & 22.418 \\ 
  h(0.72554-centroid\_max\_distance\_to\_centroid)*h(1.26151-distance\_max) & -0.718 \\ 
  h(0.28-chull\_points\_on\_hull)*h(mst\_dists\_mean-0.0671864) & 1.737 \\ 
  h(0.28-chull\_points\_on\_hull)*h(0.0671864-mst\_dists\_mean) & -15.553 \\ 
  h(distance\_mean\_tour\_length-56.3844) & 0.005 \\ 
  h(56.3844-distance\_mean\_tour\_length) & -0.002 \\ 
  h(mst\_dists\_mean-0.105423) & -0.224 \\ 
  h(0.105423-mst\_dists\_mean) & -1.681 \\ 
  h(centroid\_min\_distance\_to\_centroid-0.314769)*h(mst\_depth\_max-12) & 0.089 \\ 
  h(0.314769-centroid\_min\_distance\_to\_centroid)*h(mst\_depth\_max-12) & -0.003 \\ 
  h(angle\_mean-1.45864)*h(angle\_median-2.16409) & -0.006 \\ 
  h(angle\_mean-1.45864)*h(2.16409-angle\_median) & -0.100 \\ 
  h(angle\_mean-1.81252)*h(mst\_depth\_max-12) & -0.013 \\ 
  h(1.81252-angle\_mean)*h(mst\_depth\_max-12) & -0.003 \\ 
  h(mst\_dists\_sd-0.0369208)*h(mst\_dists\_sum-0.0178643) & 170.696 \\ 
  h(distance\_sd-0.274633)*h(12-mst\_depth\_median) & 0.075 \\ 
  h(mst\_depth\_sd-11.632) & 0.011 \\ 
  h(11.632-mst\_depth\_sd) & 0.025 \\ 
  h(mst\_depth\_mean-18.8) & 0.011 \\ 
  h(18.8-mst\_depth\_mean) & -0.017 \\ 
  h(angle\_mean-1.82054)*h(0.105423-mst\_dists\_mean) & 6.850 \\ 
  h(1.82054-angle\_mean)*h(0.105423-mst\_dists\_mean) & 0.699 \\ 
  h(distance\_max-1.26151)*h(mst\_dists\_sd-0.0495085) & 3.614 \\ 
  h(distance\_max-1.26151)*h(0.0495085-mst\_dists\_sd) & 5.079 \\ 
  h(mst\_depth\_max-12)*h(mst\_dists\_mean-0.0829785) & 0.066 \\ 
  h(mst\_depth\_max-12)*h(0.0829785-mst\_dists\_mean) & 0.082 \\ 
   \bottomrule
\end{tabular}

  \caption{Results of the MARS model for $rnd$.}
  \label{tab:MARS}
\end{table}

\begin{figure}
  \centering
  \includegraphics{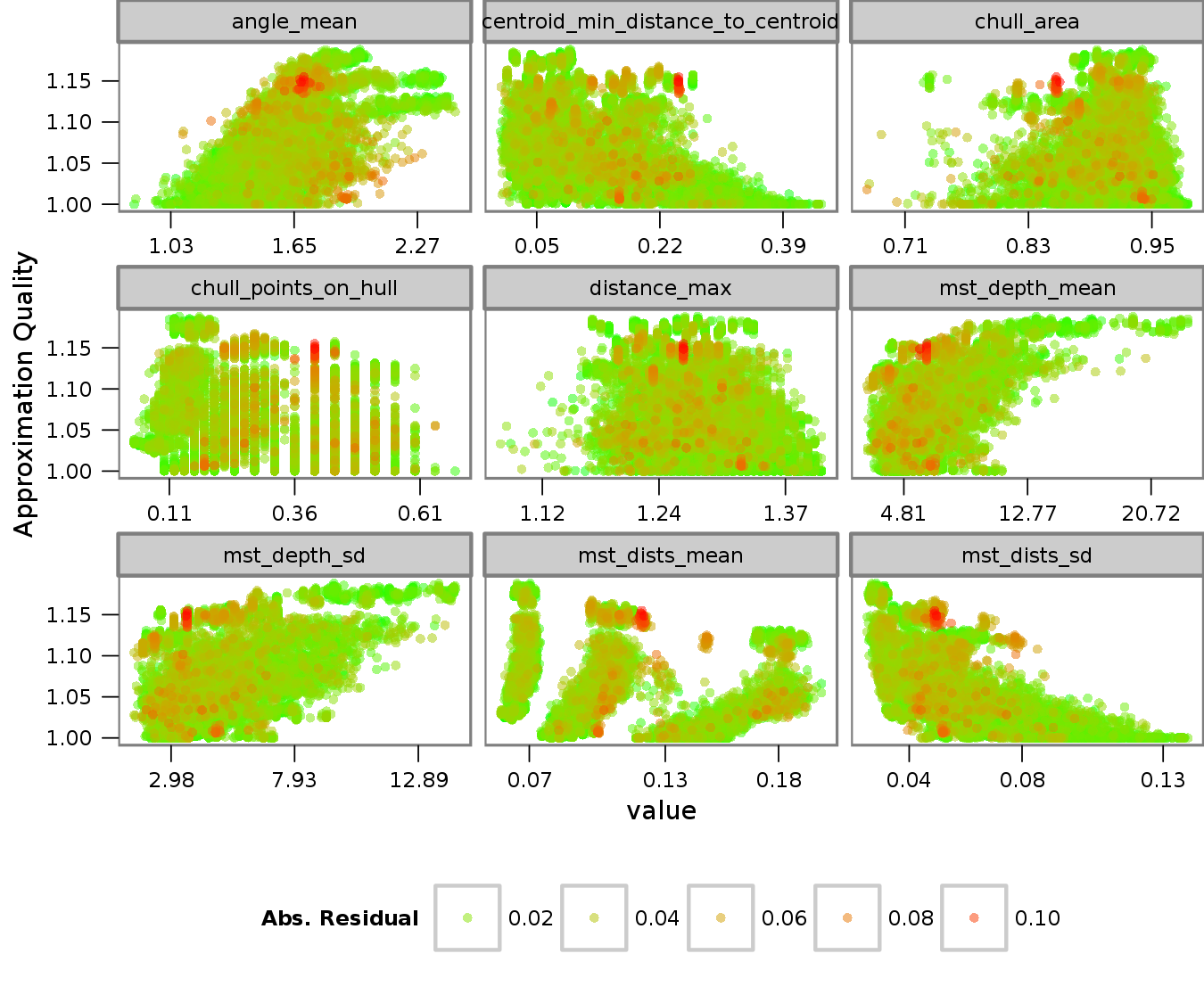}\vspace*{0.2cm}
  \includegraphics{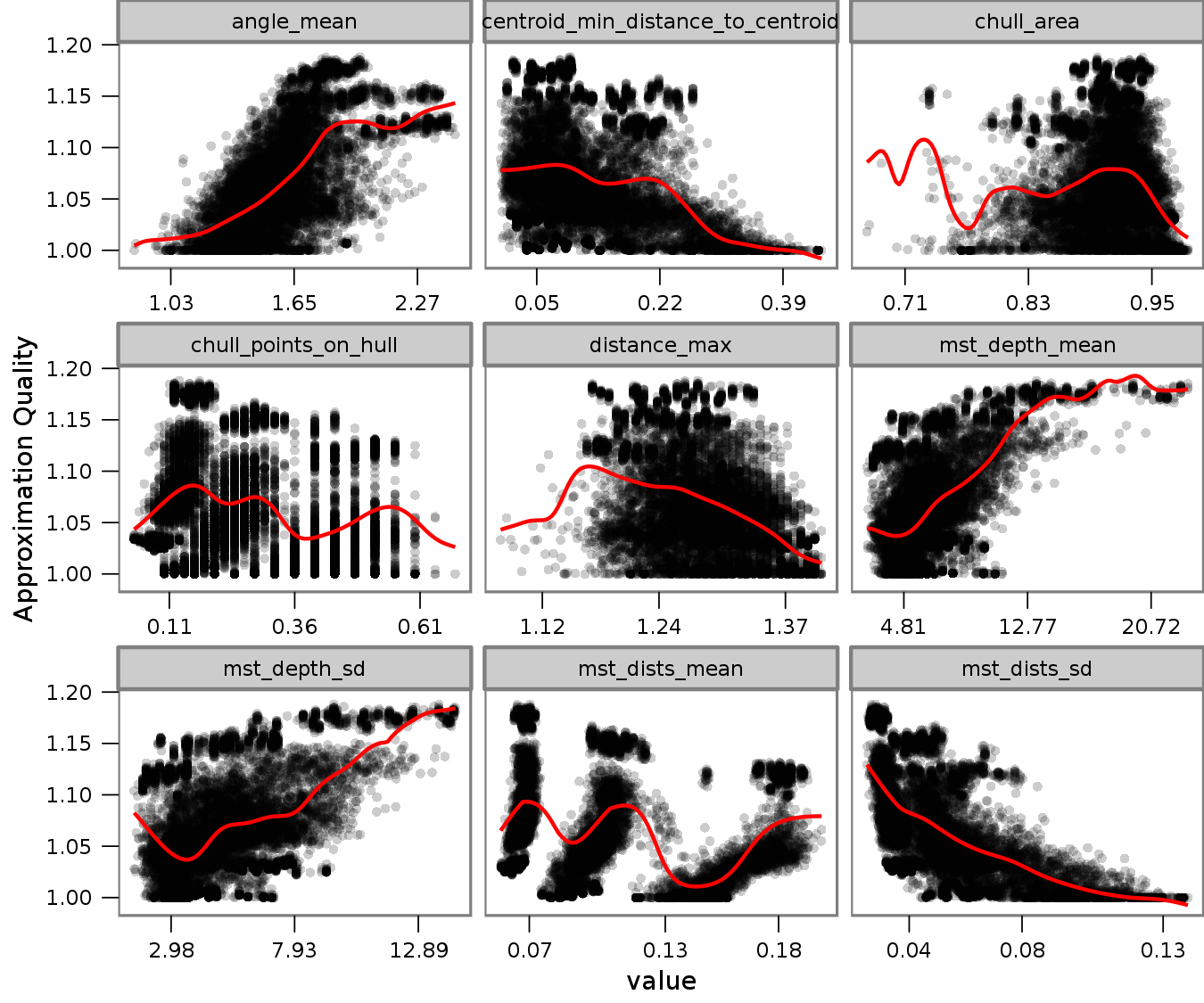}
  \caption{MARS model: Nine most frequently used features for
    $rnd$. Top: Residual scatter plot, Bottom: Weighted Partial
    Dependency Plot}
  \label{fig:MARS}
\end{figure}

Because MARS models are highly non-linear, it is hard to visualize
them. In Fig.~\ref{fig:MARS} nine features which are frequently used
in the splines of the model are visualized. The top of the figure
shows a scatter plot of the feature against the approximation
quality.

The color of each point represents the error the model makes when
predicting this point and gives us some insight into where the model
fits the data well and were it deviates significantly. The only real
structure to be seen in the plots is a small cluster of red points with an
approximation quality of about $1.15$ that is visible in every panel
of the plot. This shows that our model fits the data fairly well. The
bottom part of the figure shows a variant of a partial dependency
plot. Instead of averaging over all observations as in the partial
dependency plot, we use a weighted average, where we give observations
that are close to the feature value a higher weight - we call this the
weighted partial dependency plot. That is, for a feature $x$ with
value $x^*$, we calculate
\begin{displaymath}
  f(x^*) = \frac{1}{\sum_{i=1}^N w_i(x^*)}
  \sum_{i=1}^n w_i(x^*) * m((x^*, \vec d_i \setminus x)) \\
\end{displaymath}
where $\vec d_i$ denotes the features of the $i$-th instance,
$w_i(x^*)$ the weight assigned to the $i$-th observation and $m((x^*,
\vec d_i \setminus x))$ the predicted approximation quality for the
$i$-th feature vector if we set feature $x$ to $x^*$. We chose to use a
Gaussian weighting function
\begin{align*}
  w_i(x^*) = \alpha \phi(x_i - x^*)
\end{align*}
where $\alpha$ is set to a fourth of the standard deviation of the
feature and $\phi$ denotes the density function of the standard normal distribution. We see that the average response of the model fits the point
clouds quite well.

We further studied whether we could handle the regression problem with
a substantially smaller feature set in order to simplify our
model. For this purpose we performed a sequential forward search,
which iteratively adds the best feature w.r.t. the RMSE. As
regression model we again used MARS with second order interaction
effects. Such a forward search is a simple feature selection wrapper
as introduced in \cite{Kohavi1997}. In this procedure we perform an
outer resampling loop (here 10-fold cross-validation) to create
training and test sets. For each training set we perform the feature
selection by forward search. For each training set and feature
selection run the outer training set is resampled again (here simple
hold-out with 2/3 for training and 1/3 for testing) in a so called
inner loop. The RMSE of each feature set is measured and greedily
optimized according to this inner resampling. We stop the selection
when the performance in RMSE does not improve by at least $\sqrt{5
  \cdot 10^{-5}}$. The outer resampling ensures unbiased performance
results and the whole procedure is sometimes called nested resampling \cite{BMTW12}.

The final results are 10 potentially different feature sets, but in
our case we always end up with the four features displayed in
Table~\ref{tab:varsel}, which are also always selected in the same
order. We also display the (mean) RMSE for the feature sets during
the search on the inner test sets (the numbers are averaged across all 10
feature selections).  The unbiased RMSE in the outer cross-validation
is $0.02037$.

\begin{table}
  \centering
  \begin{tabular}{p{8cm}l}
    \toprule
    \textbf{Feature list} & \textbf{RMSE} \\
    \midrule
    empty model           & $0.05113$     \\
    + mst\_dists\_sd      & $0.03440$     \\
    + angle\_mean         & $0.02515$     \\
    + mst\_dists\_mean    & $0.02240$     \\
    + mst\_depth\_median  & $0.02036$     \\
    \bottomrule
  \end{tabular}
  \caption{Results of the MARS model with feature selection by forward search.}
  \label{tab:varsel}
  \vskip-2em
\end{table}

From the results we gain further insights into which features reduce
the RMSE the most and that we can build an acceptable model with only
four features. But it must still be noted that we perform
substantially worse than selecting the full model. The reader should
be aware of the fact that a MARS model already performs an internal
feature selection which is somewhat similar to our approach, but
faster. This last step was mainly undertaken to study in further
detail how well we can predict the approximation quality with a model
with a really low number of features.

\section{Summary and Outlook}\label{sec:conc}

In this paper we investigated concepts to predict TSP problem hardness
for $2$-opt based local search strategies on the basis of experimental
features that characterize the properties of a TSP instance. A crucial
aspect was the generation of a representative instance set as a basis
for the analysis. This turned out to be far from
straightforward. Therefore, it was only possible to generate very hard
and very easy instances using sophisticated (evolutionary) strategies.
Summarizing, we managed to generate classes of easy and hard instances
of different sizes for which we are able to predict the correct
instance class based on the corresponding feature levels with only
marginal errors.  Several feature combinations, which are cheap to
compute even for large instances, could be identified as key features
for differentiating between hard and easy instances, and the results
are supported by exploratory analysis of the evolved instances and the
respective optimal tours. However, it should be noted that most
probably not the whole space of possible hard instances is covered by
using our evolutionary method, i.e.  probably only a subset of
possible characteristics or feature combinations that make a problem
hard for $2$-opt can be identified by the applied methodology.

Instances of moderate difficulty were constructed by morphing hard
into easy instances where the effects of the transition on the
corresponding feature levels could be studied. A MARS model was
successfully applied to predict the approximation quality of $2$-opt
independent from the instance size based on the features of the
generated instances with very high accuracy. We strongly believe that
it should be straight forward to apply the same methodology to other
algorithms and use these models to derive a strategy for the algorithm
selection problem in the context of the TSP.

Moreover, we investigated two different rounding schemes within the
evolutionary algorithm for instance generation which either result in
instances exhibiting points on a regular grid or slightly perturbed
points. However, the experimental results did not show any significant
differences between the different concepts.

The analysis offers promising perspectives for further research,
specifically a systematic comparison to other local and global search
as well as hybrid solvers with respect to the influence of the feature
levels of an instance on the performance of the respective algorithms.
The investigation of much higher instances sizes would be very
interesting as well. However, it has to be kept in mind that the
computational effort intensely increases with increasing instance size
as the optimum solution, e.g. computable via Concorde, is required to
calculate the approximation quality of $2$-opt.

Finally, it is open how representative the generated instances are for
real-world TSP instances. It is therefore very desirable to collect
and create a much larger pool of small to medium sized, real-world,
TSP instances for comparison experiments. There is also the question
of how well these models can extrapolate to much larger instance
sizes. This would again be a desirable property in the context of
algorithm selection for very large instances for which it is not
feasible to calculate the global optimal tour.

In closing we would like to mention that all source code used in these experiments is available online (see Footnote \ref{fn:code}) for anyone to use and extend.


\section*{Acknowledgements}
  This work was partly supported by the Collaborative Research Center
  SFB 823, the Graduate School of Energy Efficient Production and
  Logistics and the Research Training Group ``Statistical Modelling''
  of the German Research Foundation.

\bibliographystyle{plain}
\bibliography{AMAI,book}
\end{document}